\documentclass{aa}  

\usepackage{color}
\usepackage{graphicx}
\usepackage{amsmath}% Advanced maths commands
\usepackage{amssymb}% Extra maths symbols
\usepackage{breqn}
\usepackage{gensymb}
\usepackage{placeins}
\usepackage{natbib}
\usepackage{txfonts}
\newcommand{\rg}{r_{\rm g}}
\newcommand{\tunit}{\rg/c}

\newcommand{\msun}{{M}_{\odot}}

\newcommand{\mdotu}{M_\odot {\rm yr}^{-1}}

\begin{document}

 %  \title{Cartesian AMR GRMHD simulations of black-hole accretion: \\
%the appearance of the jet-launching region in M\,87}
%
%% \title{Cartesian AMR general-relativistic magnetohydrodynamical
%%   $\kappa$-jet models of the jet-launching region in M\,87}
\title{Modeling non-thermal emission from the jet-launching region\\ of {M 87} with adaptive mesh refinement}
\titlerunning{Modeling non-thermal emission from the jet-launching region\\ of {M 87}}
\authorrunning{J. Davelaar et al.}

%\lr{Note new title}
\author{J. Davelaar \inst{1,2}\fnmsep\thanks{j.davelaar@astro.ru.nl} ~, H. Olivares \inst{3}, O. Porth \inst{4,3}, T. Bronzwaer \inst{1}, M. Janssen \inst{1}, F. Roelofs \inst{1}, \\ 
Y. Mizuno \inst{3}, C. M. Fromm \inst{3,5}, H. Falcke \inst{1}, and L. Rezzolla \inst{3}
}

\institute{
           Department of Astrophysics/IMAPP, Radboud University, P.O. Box 9010, 6500 GL  Nijmegen, The Netherlands
           \and
           Center for Computational Astrophysics, Flatiron Institute, 162 Fifth Avenue,  New York, NY 10010, USA
           \and
           Institut f\"ur Theoretische Physik, Max-von-Laue-Stra{\ss}e 1, D-60438 Frankfurt am Main, Germany
          \and
           Anton Pannekoek Instituut, Universiteit van Amsterdam P.O. Box 94249, 1090 GE Amsterdam, The Netherlands
           \and
           Max-Planck Institute for Radio Astronomy, Auf dem Huegel 69, D-53115 Bonn, Germany
          }

   \date{Received XXX ; accepted YYY}

% \abstract{}{}{}{}{} 
% 5 {} token are mandatory
 
\abstract
  % context heading (optional)
  % {} leave it empty if necessary  
  {The galaxy M\,87 harbors a kiloparsec-scale relativistic jet, whose
    origin coincides with a compact source thought to be a supermassive
    black hole. Observational mm-VLBI campaigns are capable of resolving
    the jet-launching region at the scale of the event horizon. In order
    to provide a context for interpreting these observations, realistic
    general-relativistic magnetohydrodynamical (GRMHD) models of the
    accretion flow are constructed. }
  % aims heading (mandatory)
  {Electrons in the jet are responsible for the observed synchrotron
    radiation, which is emitted in frequencies ranging from radio to
    near-infrared (NIR) and optical. The characteristics of the emitted
    radiation depend on the shape of the electrons' energy-distribution
    function (eDF). The dependency on the eDF is omitted in the modeling
    of the first Event Horizon Telescope results. In this work, we aim to
    model the M\,87 spectral-energy distribution from radio up to
    NIR/optical frequencies using a thermal-relativistic
    Maxwell-J\"uttner distribution, as well as a relativistic
    $\kappa$-distribution function. The electrons are injected based on
    sub-grid, particle-in-cell parametrizations for sub-relativistic
    reconnection.}
  % methods heading (mandatory)
   {A GRMHD simulation in Cartesian-Kerr-Schild coordinates, using eight
     levels of adaptive mesh refinement (AMR), forms the basis of our
     model. To obtain spectra and images, the GRMHD data is
     post-processed with the ray-tracing code {\tt RAPTOR}, which is
     capable of ray tracing through GRMHD simulation data that is stored
     in multi-level AMR grids. The resulting spectra and images maps are
     compared with observations. }
   % results heading (mandatory)
   {We obtain radio spectra in both the thermal-jet and $\kappa$-jet
     models consistent with radio observations. Additionally, the
     $\kappa$-jet models also recover the NIR/optical emission. The
     images show a more extended structure at 43 GHz and 86 GHz and more
     compact emission at 228 GHz. The models recover the observed source
     sizes and core shifts and obtain a jet power of
     $\approx10^{43}$~ergs/s. In the $\kappa$-jet models, both the
     accretion rates and jet powers are approximately two times lower
     than the thermal-jet model. The frequency cut-off observed at
     $\nu\approx10^{15}$ Hz is recovered when the accelerator size is
     $10^{6}-10^{8}$ cm, this could potentially point to an upper limit
     for plasmoid sizes in the jet of M 87.}
   % conclusions heading (optional), leave it empty if necessary
   %
   {}
   \maketitle
%
%________________________________________________________________

\section{Introduction}\label{sec:intro}

More than a century ago, a bright jet in the Virgo cluster was discovered
by \cite{Curtis1918}. The jet is connected to a nucleus that resides in
the center of M\,87, which is an elliptical galaxy. Since its discovery,
the jet of M\,87 has been subject to extensive radio observations
\citep{bolton1949,mills1952,baade1954,turland1975,owen2000,gasperin2012},
and various mm observations; 7 mm (43 GHz)
\citep{Junor1999,Ly2004,Walker2008,hada2011,hada2013,hada2016,walker2018},
3 mm (86 GHz) \citep{krichbaum2006,hada2013,kim2018}, and 1.3 mm (228
GHz) \citep{doeleman2012}. At mm-wavelengths, the radio emission shows a
source morphology that is consistent with a jet launched from the
putative supermassive black hole in the centre of the radio core with a
mass of $M_{\rm BH} = 6.2 \times 10^9 ~\msun$ \citep{gebhardt2011} and at
a distance of $d = 16.7$ Mpc \citep{mei2007}. This black hole is one of
the primary targets of global mm-VLBI observations by the Event Horizon
Telescope Collaboration,  which has the aim to spatially resolve black-hole
shadows \citep{falcke2000,goddi2017}, and succeeded in the case of M 87
\citep{eht-paperI,eht-paperII,eht-paperIII,eht-paperIV,eht-paperV,eht-paperVI}. The
shadow of a black hole is a depression of flux in the radiation field
surrounding the black-holes event horizon, for a non-rotating black hole
its size on the sky is given by $2\sqrt{27}G M/(c^2 D)$, with $G$ the
Gravitational constant, $M$ the black-hole mass, $c$ the speed of light,
and $D$ the distance to the black hole. Due to the large set of
observations available across the electromagnetic spectrum \citep[see
  e.g.,][]{prieto2016} and the event horizon scale mm-VLBI observations,
it is possible to use the M\,87 jet as a laboratory to study jet
launching and particle acceleration.

Since the discovery of M\,87, relativistic jets have been studied in
great detail in theory. The analytical model by \cite{blandford1979}
describes an isothermal jet model that can explain the observed flat
radio spectra of jets. They recover the observed relation between source
size ($r$) and frequency ($\nu$) to be $r\propto \nu^{-1}$. An addition
to this model was made by \cite{falcke1995}, who connected the accretion
rate to the jet.

\cite{broderick2009} modeled M\,87 with an analytic, force-free jet
model. Their best-fit model is consistent with 43 GHz observations. The
model parameters include a black-hole spin of $a_* = {J c}/{G
  M^2}=0.998$, a viewing angle of $i=25^\degree$, and a jet foot-point 
at $r=10~\rg$, where the gravitational radius $\rg$ is defined as 
$\rg = {GM}/{c^2}$. The disk consists of both thermal and accelerated 
electrons, but the fraction of accelerated electrons is low 
(around one percent). Inside the jet, only an accelerated 
electron population is present. Their model uses a black-hole mass 
of $M=3.4\times10^9 \msun$ \citep{walsh2013}.

General-relativistic magnetohydrodynamical (GRMHD) simulations are often
used to study the dynamics of accretion flows. Next, we review some of
the earlier GRMHD based models of the M\,87 jet. The first model of M\,87
based on GRMHD simulations was presented by \cite{dexter2012}, who
computed synthetic synchrotron maps based on a high-spin GRMHD
simulation. Their models included a thermal electron population in the
disk and a power-law based electron population in the jet. Their best-fit
model, at an inclination of $25^\degree$, showed counter-jet dominated
emission, meaning that most of the radiation detected by the observer
originates in the jet that points away from the
observer. \cite{dexter2012} obtained a mass-accretion rate of $\dot{M}
\approx 10^{-3} ~\msun /$yr, and a power-law index of the non-thermal 
electron distribution function of $p=3.25-3.5$, where they used a
constant electron-to-proton temperature ratio of $T_{\rm p}/T_{\rm e}=10$.

\cite{moscibrodzka2016} used GRMHD simulations and a Monte Carlo-based
radiative-transfer code to model the full spectral energy distribution
(SED) of an accreting supermassive black hole from radio to X-ray, as
well as ray-traced images of the accretion flow at $43, 86$ GHz, and
$230$ GHz. A thermal distribution function of the electrons was assumed
across the simulation domain, and the electron physics was modeled by
coupling the ion-to-electron temperature as a function of plasma $\beta =
P/P_{\rm mag}$, where $P$ is the gas pressure and $P_{\rm mag}$ the
magnetic pressure. The electrons were thermally distributed both in the
disk and the jet. The authors obtained a mass accretion rate of $\dot{M}
\approx 9\times10^{-3} ~\msun$, a favored inclination angle of
$20^\degree$ or $160^\degree$ and a ion-to-electron temperature ratio in
the disk of $T_i/T_e=100$. Smaller values of the ion-to-electron
temperature ratio resulted in an excess of X-ray emission. The 230 GHz
images showed counter-jet dominated emission. Subsequently,
\cite{moscibrodzka2017} performed polarised radiative transfer
calculations of the jet launching foot point of M\,87 to obtain Faraday
rotation measurements. It is shown that the best-fit jet-dominated model
from \cite{moscibrodzka2016} recovers the observed 1\% polarization
fraction and rotation measure of the jet base in M\,87.

Recently, \cite{ryan2018} performed 2D-axisymmetric two-temperature GRMHD
simulations that include radiative cooling. The authors conclude that
radiative cooling is important in the inner region ($r$<$10~\rg$) of the
accretion flow, and that the black-hole mass of $M=6.2\times10^{9}
~\msun$ and spin $a_* = 0.9375$ simulation recovers the observed radio
and X-ray emission and image size at 230 GHz. The jet opening angle in
their model at lower frequencies is too narrow compared to the
mm-observations of the jet base in M 87 and the model assumes a thermal
electron population in the entire simulation domain. \cite{chael2019}
also performed a two-temperature radiative GRMHD model of a Magnetically
Arrested Disc (MAD) \citep{narayan2003, tchekhovskoy2011}. The model
recovers observables such as jet opening angle, image size, core shift,
and radio SED. This model also considers a thermal electron population
and, therefore, does not fit the NIR/optical emission.

In 2019, the Event Horizon Telescope published its first set of results,
showing an asymmetric ring-like structure in the radio core of M 87 at 228 GHz. This ring-like structure is evidence for the existence of a black hole shadow and consistent with predictions from GRMHD models
\citep{eht-paperI,eht-paperII,eht-paperIII,eht-paperIV,eht-paperV,eht-paperVI}. A
detailed comparison of GRMHD models with the data can be found in
\cite{eht-paperV}. The main assumption in these models that we want to
address in this work is that the electron distribution function is taken
to be \textit{thermal} in the entire simulation domain.

All of the models have in common that they are based on GRMHD simulations
that use spherical polar coordinates with a radial grid that is
logarithmically spaced. Such a grid has the advantage of high resolution
close to the event horizon but introduces a polar axis that needs careful
treatment, potentially resulting in numerical issues that affect the jet
outflow. GRMHD codes often track only the dynamically important ion
fluid, with no direct knowledge of the electrons available. One of the
open questions in modeling the electromagnetic radiation emerging from
accreting black holes is, therefore, the shape of the distribution
function of the radiatively important electrons. The often-made
assumption that the electrons in the full simulation domain are in a
thermal-relativistic Maxwell-J\"uttner distribution potentially breaks
down in regions where non-ideal effects are important.

\begin{figure}
\centering
    \includegraphics[width=0.45\textwidth]{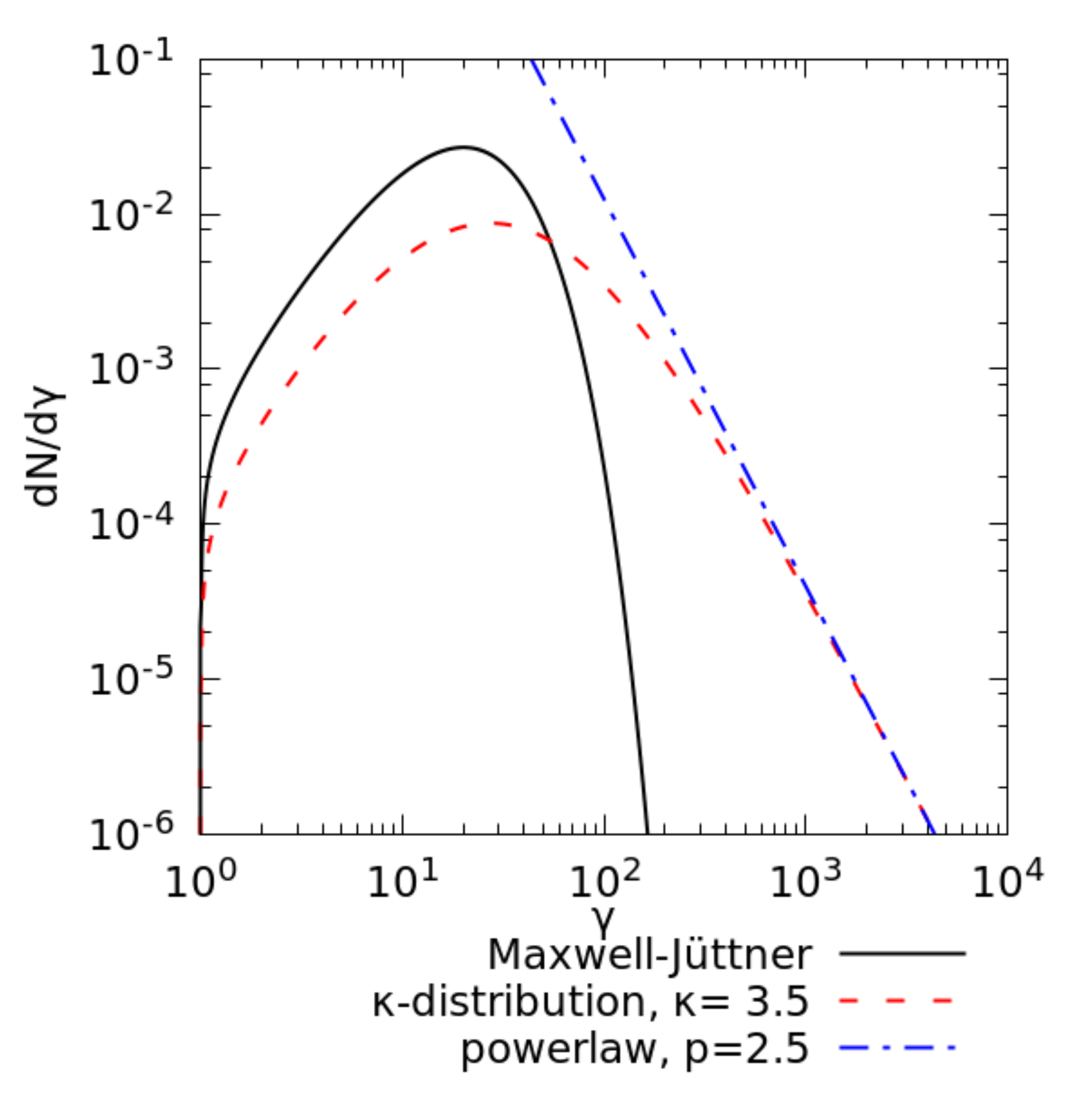}
       \caption{Maxwell-J\"uttner distribution (black),
         $\kappa$-distribution (orange) for a $\kappa$ value of 3.5, and
         a power-law distribution (yellow) with $p=2.5$.}
        \label{fig:distr}
\end{figure}

These non-ideal effects are expected to be strongest in the highly
magnetized regions of the jet, where they can be associated with magnetic
reconnection accelerating electrons to very large energies. In the case
of M\,87, features of electron acceleration are observed in the
NIR/optical wavebands \citep[see e.g.,][and references
  therein]{prieto2016}. We, therefore, need a distribution function that
describes the electrons that are not in thermal equilibrium. Particle
ensembles that are not in thermal equilibrium {can be} described in the
framework of Tsallis statistical mechanics \citep{Tsallis1988}. In this
framework, the $\kappa$-distribution function plays a key role. In
Fig.~\ref{fig:distr} we show that the $\kappa$-distribution function is a
combination of a thermal core at low values of the Lorentz factor
$\gamma$, which asymptotically turns into to a power-law with power-law
index $p=\kappa-1$ for large $\gamma$ values. In the limit of $\kappa
\rightarrow \infty$, the $\kappa$-distribution becomes the
Maxwell-J\"uttner distribution function \citep{Rezzolla_book:2013}. The
$\kappa$-distribution function is observed at a variety of astrophysical
systems such as the solar wind, solar magnetosphere, Jovian
magnetospheres, planetary nebula, and many others (see for a review
\cite{pierrard2010} and references therein).

In \cite{davelaar2018}, we introduced a $\kappa$-jet model for the
accreting black hole in the center of the Milky Way, Sagittarius A* (Sgr
A*). This model is a combination of a thermal and a $\kappa$-distributed
electron population. In the accretion disk, we inject thermal electrons,
while in the jet we inject a mix of thermal and $\kappa$-distributed
electrons. The ratio between the two species is a free parameter of the
model. In the case of Sgr A*, we found that $\sim 5-10\%$ of 
the electrons is $\kappa$-distributed in the event of flares, and they
are negligible in the quiescent state. The injection method used in this
model was a uniform injection in the outflow of the simulation domain
with a fixed power-law index.

To improve the model we here connect the electron-acceleration parameters
to information from local kinetic plasma simulations. Kinetic plasma, or
particle-in-cell (PIC) simulations, are capable of resolving the
micro-physics scales that GRMHD simulations cannot reach. Although local,
these type of simulations can provide first-principle parametrizations of
particle-acceleration processes. For our model, we consider a
parametrization of the power-law index for trans-relativistic
reconnection as found by \cite{ball2018}. Reconnection is known to be an
efficient particle accelerator in magnetized environments \citep[see
  e.g.,][]{sironi2014,guo2014,sironi2015,werner2016,petropoulou2016,werner2017,werner2018}. Besides
this parametrization,we also extended our model with an injection radius,
which corresponds to the footpoint of the jet where electron acceleration
can become important.

In this work, we apply thermal and $\kappa$-jet models to the accreting
black hole in M\,87. The dynamics of the accretion flow are drawn from
GRMHD simulations performed in Cartesian-Kerr-Schild (CKS)
coordinates. {This prevents numerical artefacts and directional biases of
  the jet caused by the presence of a polar axis, this will be studied in
  detail a future work}. In addition, the use of adaptive mesh refinement
(AMR) allows us to capture the instabilities in the jet sheath, and, at
the same time, to resolve the magneto-rotational instability (MRI) in the
disk. We use the results of this simulation to generate SEDs, synthetic
synchrotron maps (images), and optical-depth maps of the jet-launching
region in M\,87. We extend the general-relativistic-ray-tracing (GRRT)
code {\tt RAPTOR}, rendering it compatible with AMR data structures. We
fit synthetic SEDs obtained from our GRRT simulations to observational
data.

The plan of the paper is as follows: in Section \ref{sec:methods} we
describe our GRMHD simulation setup, as well as the electron model
that we use in our radiative-transfer calculations. In Section
\ref{sec:results} we compute SEDs, synchrotron and opacity maps, source
sizes, and core shifts. In Section \ref{sec:disc}, we compare our results
to previous works and observations. We summarize our results in Section
\ref{sec:conc}.

\section{Methods}\label{sec:methods}

In this Section, we describe the GRMHD simulation setup, the coordinates
used to simulate the accretion flow and radiation transport, and
introduce our electron-physics model.

\subsection{GRMHD simulations}
The dynamics of the accretion flow onto the black hole are simulated
using the Black Hole Accretion Code \citep[\texttt{BHAC},][]{porth2017},
which solves the GRMHD equations
\begin{subequations}
\begin{align}
\nabla_{\mu} (\rho u^{\mu}) &= 0 \,, \\
\nabla_{\mu} T^{{\mu \nu}} &= 0 \,,  \\
\nabla_{\mu}\, ^{*}\!F^{\mu\nu} &= 0 \,, \label{eq:grmhd}
\end{align}
\end{subequations}
where $\nabla_{\mu}$ denotes the covariant derivative, $\rho$ the
rest-mass density, $u^\mu$ the fluid 4-velocity, $T^{{\mu \nu}}$ the
energy-momentum tensor of the combined perfect fluid and electromagnetic
fields, and $^{*}\!F^{\mu\nu}$ the dual of the Faraday tensor
($F^{\alpha\beta}$).

The system is closed by the ideal-MHD approximation corresponding to a
plasma with infinite conductivity $F^{\mu\nu}u_\nu =0$, and by the
equation of state of an ideal fluid \citep[see,
  e.g.,][]{Rezzolla_book:2013} $h(\rho,P)= 1
+\frac{\hat{\gamma}}{\hat{\gamma}-1}\frac{P}{\rho}$, where $h$ and $P$
are the specific enthalpy and gas pressure in the fluid frame, and the
adiabatic index $\hat{\gamma}=4/3$. The simulation is initialized with a
Fishbone-Moncrief torus \citep{fm1976} with its inner radius at $6~\rg$
and its pressure maximum at $12~\rg$. The initial configuration of the
magnetic field is a single poloidal loop described by the vector
potential $A_\phi \propto \max(\rho/\rho_{\rm max} - 0.2,0)$. The
initial density and pressure are normalized so that $\rho_{\rm max} =
1$. The initial magnetic field is also normalized such that the ratio
between maximum gas pressure $P_{\rm max}$ and maximum magnetic pressure
$P_{\rm mag,max}$ is ${P_{\rm max}}/{P_{\rm mag,max}} = 100$. The disk
is, therefore, weakly magnetized. In order to break the initial
equilibrium state and accelerate the development of the MRI, we add 5\%
`white noise' random perturbations to the pressure. This triggers the
MRI, which transports angular momentum and allows accretion onto the
black hole \citep{Balbus1991}.

The black-hole's dimensionless spin parameter was set to be $a_* =
0.9375$, where $J$ is the angular momentum. For this value of $a_*$ the
inner horizon is at $r\approx 1.34799 ~\rg$.

\subsection{AMR grid in Cartesian-Kerr-Schild coordinates}

\begin{table}
\caption{Maximum AMR refinement radii in $\rg$ for the different AMR
  levels. The jet region is defined as the region where
  $\theta<15^\degree$ or $\theta>165^\degree$, where the disk region is
  $15^\degree < \theta < 165^\degree$. }
  \label{tab:amrlevels}
\centering
\begin{tabular}{ccc} 
 \hline
 \hline
 Level  & jet &  disk \\
 \hline
 8 & $2$    & $2$  \\
 7 & $50$   & $22$ \\
 6 & $100$  & $25$ \\
 5 & $150$  & $40$ \\
 4 & $200$  & $100$ \\
 3 & $400$  & $200$ \\
 2 & $800$  & $400$ \\
 1 & $1000$ & $1000$ \\
 \hline
 \hline
\end{tabular}
\end{table}

The simulation is performed on a Cartesian (rectangular) grid. The
covariant metric $g_{\mu\nu}$ of a rotating black hole in
Cartesian-Kerr-Schild (CKS) coordinates is given by \citep[see,
  e.g.,][]{Rezzolla_book:2013}
\begin{equation}
g_{\mu\nu} = \eta_{\mu\nu} + f l_\mu l_\nu,
\end{equation}
where $\eta_{\mu\nu}={(-1,1,1,1)}$ is the Minkowski metric, and
\begin{subequations}
\begin{align}
f &= \frac{2r^3}{r^4 + a^2 z^2},\\
l_\nu &= \left(1, \frac{rx+ay}{r^2 + a^2}, \frac{ry-ax}{r^2 + a^2}, \frac{z}{r}\right),
\end{align}
\end{subequations}
where $r$ is given by
\begin{equation}
r^2 = {\frac{R^2 - a^2 + \sqrt{(R^2 - a^2)^2 + 4a^2z^2}}{2} },
\end{equation}
and 
\begin{eqnarray}
R^2 = x^2+y^2+z^2.
\end{eqnarray}
All units of length are scaled by the gravitational radius $\rg$ which is
given by $\rg = {GM}/{c^2}$. In the limit of $r \gg 0$, the radius $r
\rightarrow R$. The contravariant metric is given by
\begin{equation}
g^{\mu\nu} = \eta^{\mu\nu} - f l^\mu l^\nu,
\end{equation}
where $l^\nu$ is given by
\begin{eqnarray}
l^\nu = \left(-1, \frac{rx+ay}{r^2 + a^2}, \frac{ry-ax}{r^2 + a^2}, \frac{z}{r}\right).
\end{eqnarray}

The use of AMR allows us to increase the resolution in regions of
interest during runtime. The decision to refine is made based on the
L\"ohner scheme \citep{Loehner87}, which quantifies variations of the
density and the plasma magnetization $\sigma$, defined as
$\sigma=b^2/\rho$, where $b=\sqrt{b^{\mu}b_{\mu}}$ is the magnetic-field
strength in the fluid frame. The code is allowed to refine up to a
maximum level of refinement that depends on the location in the
computational domain; greater levels of refinement are allowed in the
regions where the jet is expected to form and the disk is expected to
reside. This distinction is made based on a radius $r$ and polar angle
$\theta$, and for the jet this region is between $\theta<15^\degree$ or
$\theta>165^\degree$. The maximum allowed refinement level as a function
of radius and polar angle are shown in Table \ref{tab:amrlevels}. The
base resolution of the grid is $96 \times 96 \times 192$ cells in $x$,
$y$, and $z$-directions, respectively. The simulation domain is $x \in
(-500 ~\rg,500 ~\rg)$, $y \in (-500 ~\rg,500 ~\rg)$ and $z \in (-1000
~\rg,1000 ~\rg)$. We simulate up to $t_{\rm f}=10^4 ~\tunit$, which
corresponds to $37.5$ orbital periods of the accretion disk at the
pressure maximum. At the end of the simulation, the domain contains
around 70 million cells.

\subsection{Ray tracing in AMR CKS grid}

In order to perform general-relativistic ray-tracing calculations in
Cartesian coordinates within the block-based AMR data structure of BHAC,
it has been necessary to extend our general-relativistic ray-tracing code
{\tt RAPTOR} \citep{bronzwaer2018}. In particular, the initial conditions
for the rays, also called the ``virtual camera'', employ a tetrad basis
in which the initial wave-vectors are described \citep{noble2007}, a
description of the implementation of this in {\tt RAPTOR} can be found in
\cite{davelaar2018b}. The tetrad camera uses a set of trial vectors to
generate a tetrad basis by using a Gramm-Schmidt orthogonalization
procedure. In spherical coordinate systems, the trial vectors are unit
vectors pointing along the $t,r,\theta,\phi$-directions. In our case, we
have to transform this into Cartesian coordinates. The virtual camera is
constructed at a position ($x_{\rm c},y_{\rm c},z_{\rm c}$) in space
which is computed based on the following parameters: \textit{(i)} the
radial distance between the camera and the black hole $r_{\rm c}$;
\textit{(ii)} the inclination with respect to the black hole spin axis
$\theta_{\rm c}$; \textit{(iii)} the azimuthal angle around the black
hole spin axis $\phi_{\rm c}$. The tetrad trial vectors can then be
defined as
\begin{subequations}
   \begin{align}
    t^\mu_{0} &= (1 ,0 ,0 ,0), \\
    t^\mu_{1} &= (0, -\sin(\theta_{\rm c})\cos(\phi_{\rm c}), -\sin(\theta_{\rm c})\sin(\phi_{\rm c}),-\cos(\theta_{\rm c})),\\
    t^\mu_{2} &= (0, -\sin(\phi_{\rm c}),\cos(\phi_{\rm c})),\\
    t^\mu_{3} &= (0, -\cos(\theta_{\rm c})\cos(\phi_{\rm c}), -\cos(\theta_{\rm c})\sin(\phi_{\rm c}),-\sin(\theta_{\rm c})).
\end{align}
\end{subequations}

The choice of trial vectors results in a right-handed basis where the
observer is facing the black hole.

The integration of the geodesic equations is done by solving the
second-order differential equation
\begin{equation}
\frac{{\rm d}^2 x^{\alpha}}{{\rm d}\lambda^2} = -\Gamma^{\alpha}_{\ \mu
  \nu} \frac{{\rm d}x^{\mu}}{{\rm d}\lambda} \frac{{\rm d}x^{\nu}}{{\rm
    d}\lambda}.
\end{equation}
where $\Gamma^{\alpha}_{\ \mu \nu}$ are the connection coefficients,
$x^{\alpha}$ is the geodesic position, and $\lambda$ is the affine
parameter. We use a fourth-order Runge-Kutta algorithm, where the
connection coefficients are evaluated using a finite-difference
derivative of the metric.

The step-sizing for the geodesic integration in {\tt RAPTOR} was adopted
since it relies on spherical logarithmic coordinates. First we compute a
required step-size based on the geodesic wave-vector 
\begin{subequations}
\begin{align}
{\rm d} \lambda_{x} &= \Delta \ / \left( \left| k^x \right| + \delta \right), \\
{\rm d} \lambda_{y} &= \Delta \ / \left( \left| k^y \right| + \delta \right), \\
{\rm d} \lambda_{z} &= \Delta \ / \left( \left| k^z \right| + \delta \right), \\
%r &= \sqrt{x^2 + y^2 + z^2},\\
%{\rm d} \lambda_{\rm geod}&=\frac{r}{{\left| {\rm d}\lambda_{x} \right|}^{-1} + {\left| {\rm d}\lambda_{y}^{-1} \right|} + {\left| {\rm d}\lambda_{z} \right|}^{-1}},
{\rm d} \lambda_{\rm geod}&=\frac{R}{{\left| {\rm d}\lambda_{x} \right|}^{-1} + {\left| {\rm d}\lambda_{y}\right|}^{-1} + {\left| {\rm d}\lambda_{z} \right|}^{-1}},
\end{align}
\end{subequations}
where $k^{x,y,z}$ are the wave-vector components in the $x,y,z$
directions, $\delta$ is a small real number to prevent divisions by zero,
and $\Delta$ is a scale factor for the step-size (typically $\Delta\approx0.01$.
Then we compute a required step-size based
the AMR cell size ${\rm d}x$
\begin{subequations}
\begin{align}
k_{\rm max} &= \max(k^x ,\max(k^y,k^z)),\\
{\rm d} \lambda_{\rm grid} &= \frac{{\rm d}x}{n k_{\rm max}},
\end{align}
\end{subequations}
where $n$ sets the amount of steps per cell. We typically use at least
two steps per cell. We then compare both the geodesic and AMR based
step-sizes and use the smallest of the two to ensure convergence; ${\rm d}
\lambda = \min({\rm d} \lambda_{\rm geod},{\rm d} \lambda_{\rm grid})$.

For the radiative-transfer part of the ray-tracing calculation, we need
the plasma variables at the location of the geodesics. We interfaced {\tt
  RAPTOR} with the AMR data structure of {\tt BHAC}, and reconstruct the
full AMR grid. The {\tt BHAC} AMR block-based data structure is parsed by
the code. When we integrate the geodesics we use a nearest-neighbor
approach to interpolate the grid-based plasma variables to the geodesics.

\subsection{Electron model and radiative-transfer model parameters}

Since GRMHD simulations are scale-free, we have to re-scale the plasma
variables from code units to c.g.s. units. Units of length are scaled
with ${\mathcal L}=\rg$, while units of time are scaled with ${\mathcal
  T} = \tunit$, the mass unit is set by ${\mathcal
  M}=1.8\times10^{29}\,{\rm gram}$. Estimates of the mass of M\,87 are used to
constrain the length and time units, we use a mass of
$M=6.2\times10^9~\msun$ \citep{gebhardt2011}, the mass used in 
this work is slightly smaller than the mass of $M=(6.5\pm0.7)\times10^9~\msun$
reported in \cite{eht-paperI}, but the used value for the black hole mass is within the error margins.
The mass unit $\mathcal{M}$, which sets the accretion rate,
however, is unknown. It is, therefore, a fit parameter. The mass unit is
directly proportional to the accretion rate via $\dot{M}_{\rm cgs} =
\dot{M}_{\rm sim} {\mathcal M} {\mathcal T}^{-1}$, where $\dot{M}_{\rm
  sim}$ is the accretion rate in simulation units. In order to scale the
relevant plasma quantities to c.g.s units, the following scaling
operations are performed: $\rho_0 = \mathcal{M}/\mathcal{L}^3$,
$u_{0}=\rho_0 c^2$, and $B_0=c\sqrt{4\pi \rho_0}$.
  
As mentioned before, our GRMHD simulation only simulates the dynamically
important protons. Therefore, we need to parametrize the electron
properties, such as their distribution functions, densities, and
temperatures, in post-processing. The plasma is assumed to be
charge-neutral, so that $n_{\rm e} = n_{\rm p}$ throughout the
domain. For the electron temperature we employ the parametrization of
\cite{moscibrodzka2016}:
\begin{subequations}
\begin{align}
T_{\rm ratio}&= T_{\rm p}/T_{\rm e} = R_{\rm low}\frac{1}{1+\beta^2} + R_{\rm high} \frac{\beta^2}{1+\beta^2},\\
\Theta_{\rm e} &= \frac{U(\hat{\gamma} - 1) m_{\rm p}/m_{\rm e}}{\rho T_{\rm ratio}},
\end{align}
\end{subequations}
where $m_{\rm p}$ is the proton mass, $m_{\rm e}$ is the electron mass,
$U$ is the internal energy, $\Theta_{\rm e}$ is the dimensionless
electron temperature that can be re-scaled to c.g.s units via $T =
{\Theta_{\rm e} m_{\rm e} c^2}/{k_{\rm b}} $, where $k_{\rm b}$ is the
Boltzmann constant. The parameters $R_{\rm low}$ and $R_{\rm high}$ are
free parameters of the model; $R_{\rm low}$ sets the temperature ratio in
the jet, where $\beta \ll 1$, and $R_{\rm high}$ sets the temperature
ratio in the disk where $\beta \gg 1$.

For the electrons' energy-distribution function, we follow a similar
recipe as described in \cite{davelaar2018}. we use the relativistic
{ isotropic} $\kappa$-distribution function for the
electrons, which is given by \citep{xiao2006}
\begin{equation}
\frac{dn_{\rm e}}{d\gamma} = N \gamma \sqrt{\gamma^2 -1} \left( 1 +
\frac{\gamma -1}{\kappa w}\right)^{-(\kappa + 1)},
\end{equation}
where $\gamma$ is the Lorentz factor of the electrons,
$\kappa$ is the parameter that sets the
power-law index $p$ via $p=\kappa-1$, $w$ sets the width of the
distribution function, and $N$ is a normalization factor such that 
the electron distribution function contains $n_e$ electrons.

The width $w$ of the $\kappa$ distribution sets the amount of energy in
the distribution. In the case that $\kappa w \gg 1$ the total energy in
the $\kappa$ distribution is given by
\begin{equation}
    E_{\kappa}  = \frac{3 \kappa}{\kappa -3} n_{\rm e} w.
\end{equation}

We couple this energy to the energy present in a thermal distribution
($E_{\rm thermal} = 3 n_{\rm e} \Theta_{\rm e}$) and add a source term
based on the magnetic energy
\begin{equation}
    E_{\kappa} = \frac{3 \kappa}{\kappa -3} n_{\rm e} w = 3 n_{\rm e}
    \Theta_{\rm e} + \tilde{\epsilon} \frac{B^2}{8\pi},
\end{equation}
here $\tilde{\epsilon}$ is used to join smoothly between between the
$\kappa$-distribution and the magnetic energy. After a bit of algebra, we
can rewrite the width as
\begin{equation}
    w = \frac{ \kappa -3 }{\kappa} \Theta_{\rm e} + \tilde{\epsilon}
    \frac{ \kappa -3 }{6 \kappa} \frac{m_{\rm p}}{m_{\rm e}} \sigma.
\end{equation}

In the limit of $\sigma \ll 1$, the $\kappa$-distribution energy is set
by the thermal energy, while in the magnetized regime the energy is set
by the magnetic energy. The $\tilde{\epsilon}$ parameter is set by
\begin{equation}
    \tilde{\epsilon} = \epsilon \frac{1}{2}\left(1 + \tanh( r - r_{\rm inj} )\right).
\end{equation}
where $r_{\rm inj}$ is the injection radius from which we start injecting
electron based on the magnetic energy, and $\epsilon$ is the base value
for radii larger than $r_{\rm inj}$; hereafter, we will consider two 
cases: where $\epsilon$ is zero or non-zero. 

The power-law index of the electrons distribution functions (eDFs) is
based on sub-grid particle-in-cell (PIC) simulations of
trans-relativistic reconnection by \cite{ball2018}, who simulated
two-dimensional reconnection layer (Harris sheath) for an electron-ion
plasma for multiple values of the plasma $\beta$ and of the magnetization
$\sigma$. One of the benefits of this type of plasma simulation is that
one obtains eDFs from first principles. In \cite{ball2018} these eDFs are
then used to fit the power-law index $p$ as a function of $\beta$ and
$\sigma$ as
\begin{subequations}
   \begin{align}
    p &= A_p + B_p \tanh\left(C_p \beta\right)\\
    A_p &= 1.8 + 0.7/ \sqrt{\sigma}\\
    B_p &= 3.7\, \sigma^{-0.19}\\
    C_p &= 23.4\, \sigma^{0.26}
    \end{align}
\end{subequations}
These fits are obtained for $10^{-4}<\beta<1.5$ and $0.1<\sigma<7.2$,
which corresponds to the typical values that we find in the jet sheath,
which is the main source of synchrotron emission in our jet-models.

\begin{figure}
\centering
    \includegraphics[width=0.45\textwidth]{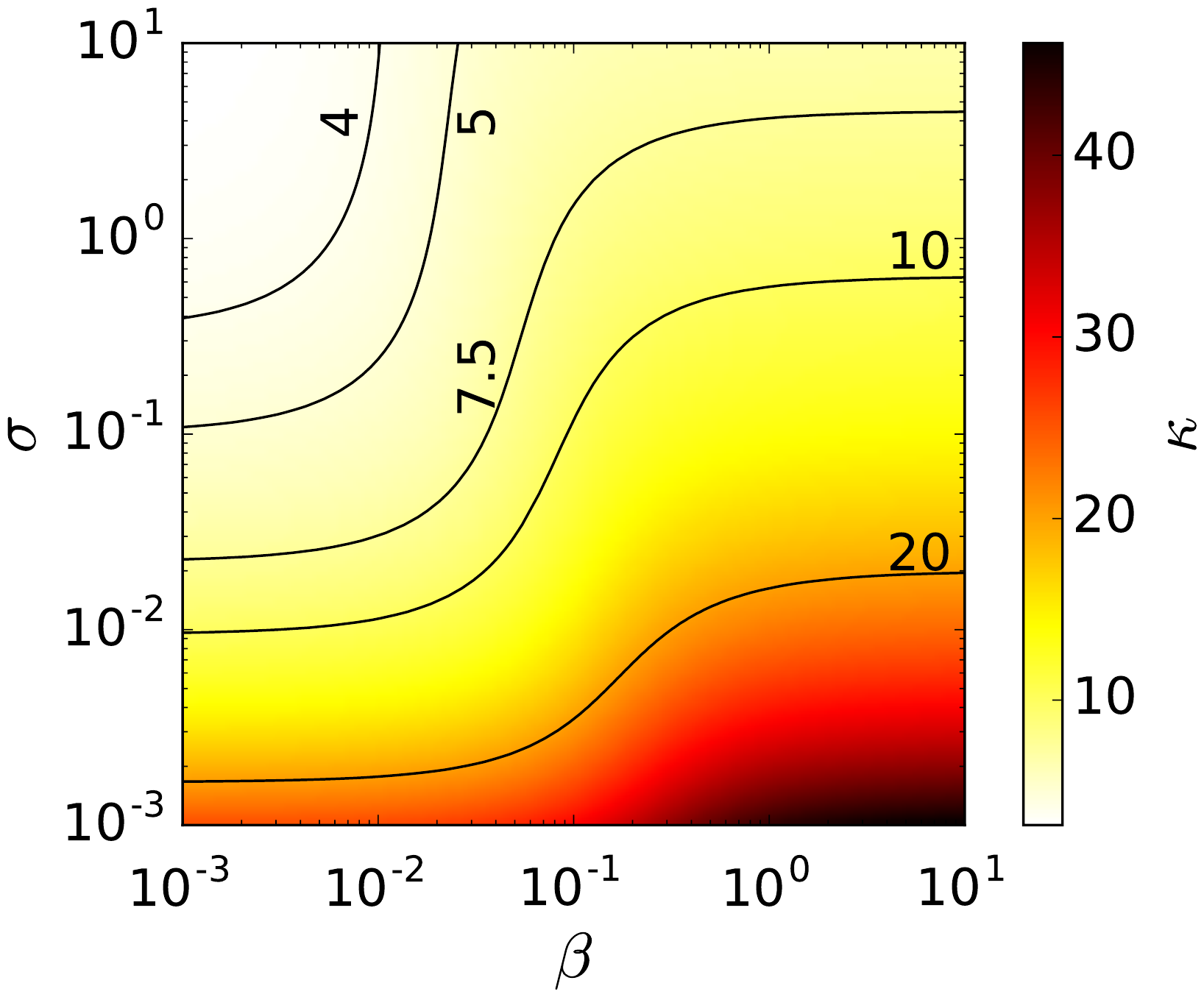}
       \caption{The $\kappa$ parameter as function of $\beta$ and
         $\sigma$ from the parameterisation as found by
         \cite{ball2018}. A high value of $\kappa$ corresponds to steep
         particle spectra with power-law index $p=\kappa-1$. We
         overplotted contours of constant $\kappa$ in black. }
        \label{fig:injection}
\end{figure} 

Finally, we exclude all emission from regions where $\sigma>5.0$, this is
what we call the jet spine. These regions are unreliable for modeling
since the thermal energy in highly magnetized regions is unreliable in
GRMHD simulations. We also exclude all emission from regions where floor
values are applied, {these regions typically resides inside the
  magnetized jet}. This results in three regions inside our simulation
domain; the disk, the jet sheath, and jet spine. The disk resides where
$\sigma$ is much smaller than one and plasma $\beta$ is large than one,
the jet sheath resides where $\sigma$ is of the order unity and $\beta$
is smaller than one. In the case of our $\kappa$-jet model we set the
electron distribution function to a relativistic $\kappa$-distribution
function into the disk and jet sheath and no
electrons are present in the jet spine.

The emission and absorption coefficients for the thermal electron
distributions are taken from \cite{leung2011}, and in the case of the
$\kappa$-distribution, the fit formula taken are from \cite{pandya2016}.

\subsection{SED cut off}

The SED of M\,87 shows a clear cut-off in flux around $\nu = 10^{15}$ Hz
\citep{prieto2016}. We will consider three potential sources for this
cut-off.

First, we assume that the cut-off is caused by synchrotron cooling in
the jet, which becomes important when the synchrotron-cooling time of the
electron is comparable with the typical dynamical time. Under these
conditions, the cooling (cut-off) frequency is given by
\begin{equation}
    \nu_{\rm cool} = \frac{18 \pi}{\sigma_T^2} \frac{m_e c^2 e }{B^3 z_{\rm jet}},
\end{equation}
where $\sigma_T$ is the Thomson cross-section, and $z_{\rm jet}$ the
position along the jet.

Second, we assume that the break takes place at the synchrotron burn-off
limit, that is, at the maximum energy that a particle can gain while
emitting synchrotron radiation. The maximum Lorentz factor in this case
is
\begin{equation}
\gamma_{\rm max} = \sqrt{\frac{3m_{\rm e}^2 c^4 E}{4\pi  e^3 B^2}},
\end{equation}
where $E$ is the electric field, and the cut-off frequency is then given
by
\begin{equation}\label{eq:cut-off}
  \nu_{\rm cut-off} = \frac{3}{2} \gamma_{\rm max}^2 \nu_c,
\end{equation}
with $\nu_c = {eB}/(2 \pi m_{\rm e} c)$.
 
Finally, we assume that break is given by the Hillas criterion \citep{hillas1984},
stating that the maximum Lorentz factor achievable can be estimated by equating the
gyration radius of the electron and the size of the acceleration region
$L$. This results in a maximum Lorentz factor of
\begin{equation}
  \label{eq:hillas}
\gamma_{\rm max} = \frac{eBL}{m_{\rm e} c^2},
\end{equation}
which results in a cut-off frequency of $\nu\approx10^{15}$ Hz after
using Eq. \eqref{eq:hillas} in \eqref{eq:cut-off}. In this way, we can
also we can estimate the typical size $L$ of the acceleration-region 
\begin{equation}
  \label{eq:lengthscale}
  L =  \sqrt{\frac{4\pi \nu_{\rm cut-off} m_{\rm e}^3 c^5}{e^3 B^3}} \approx 4.5 \times 10^{7} {\rm cm} \sqrt{\frac{(\nu_{\rm cut-off}/10^{15} {\rm Hz})}{{(B/1 ~{\rm G})^3}}}.
\end{equation}
 Interestingly, the maximum size $L$ can be interpreted as the
size of plasmoids as was done by \cite{petropoulou2016} and
\cite{christie2019} for blazars.

\section{Results}\label{sec:results}

In this Section, we present the results of our GRMHD simulations and how
the SEDs they produce can be compared with the observational data.  at
three observational relevant frequencies at two inclinations. We also
show how we can compute from the synthetic images the source size and
core shifts, and how they compare with the observations.

\subsection{Structure of the accretion disk and jet in the AMR simulation}

\begin{figure*}
    \centering
    \includegraphics[width=0.9\textwidth]{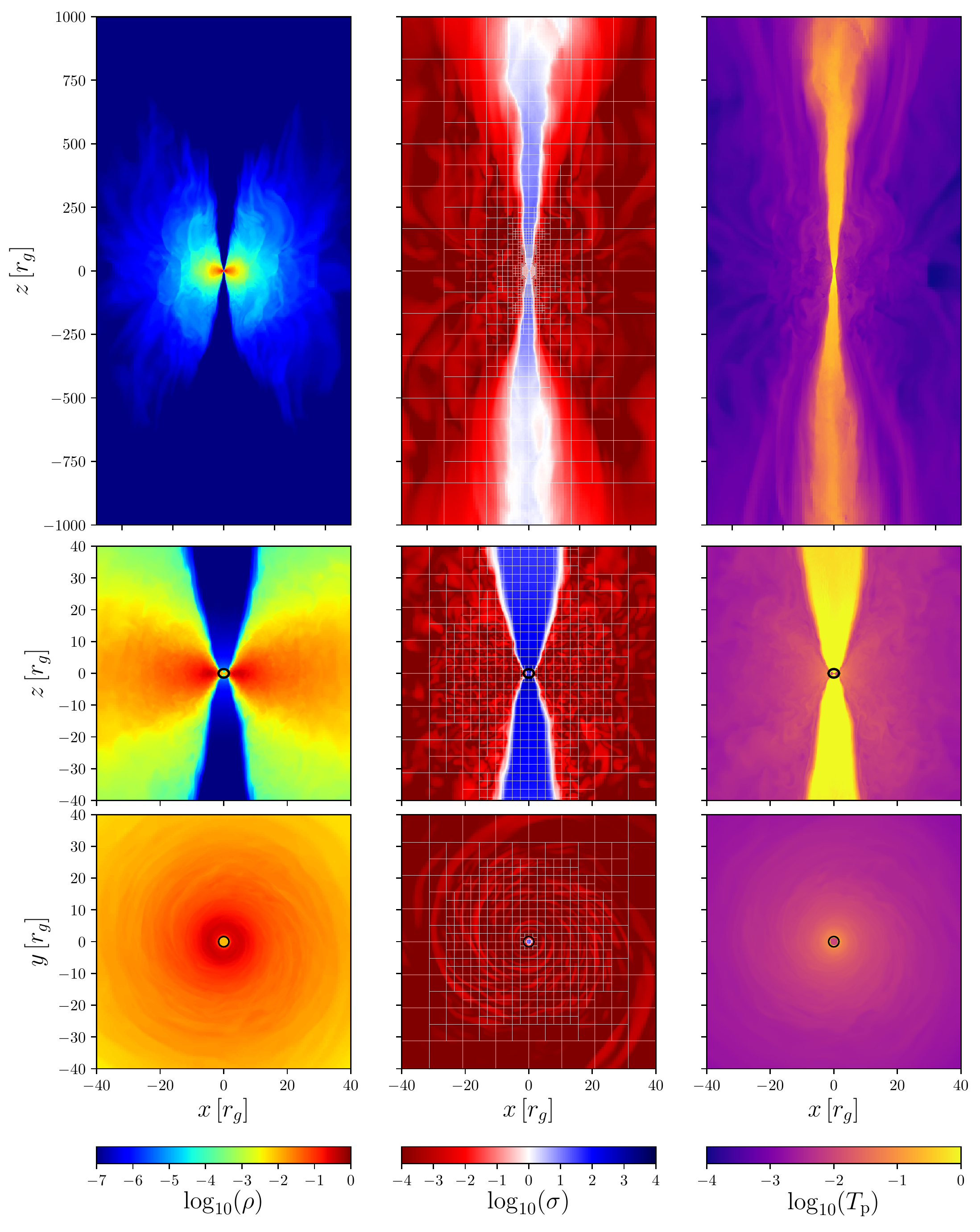}
        \caption{Left panels: slice in the $(x,z)$ and $(x,y)$ planes of
          the density in code units. Middle panels: slice along the
          $(x,z)$ and $(x,y)$ planes of the magnetization parameter
          $b^2/\rho$, over-plotted with the grid block sizes. Right
          panels: slice along the $(x,z)$ and $(x,y)$ planes of the
          dimensionless ion temperature. Shown with a black circle is the
          location of the event horizon. }
        \label{fig:snapshots}
\end{figure*} 

\begin{figure*}
    \includegraphics[width=0.49\textwidth]{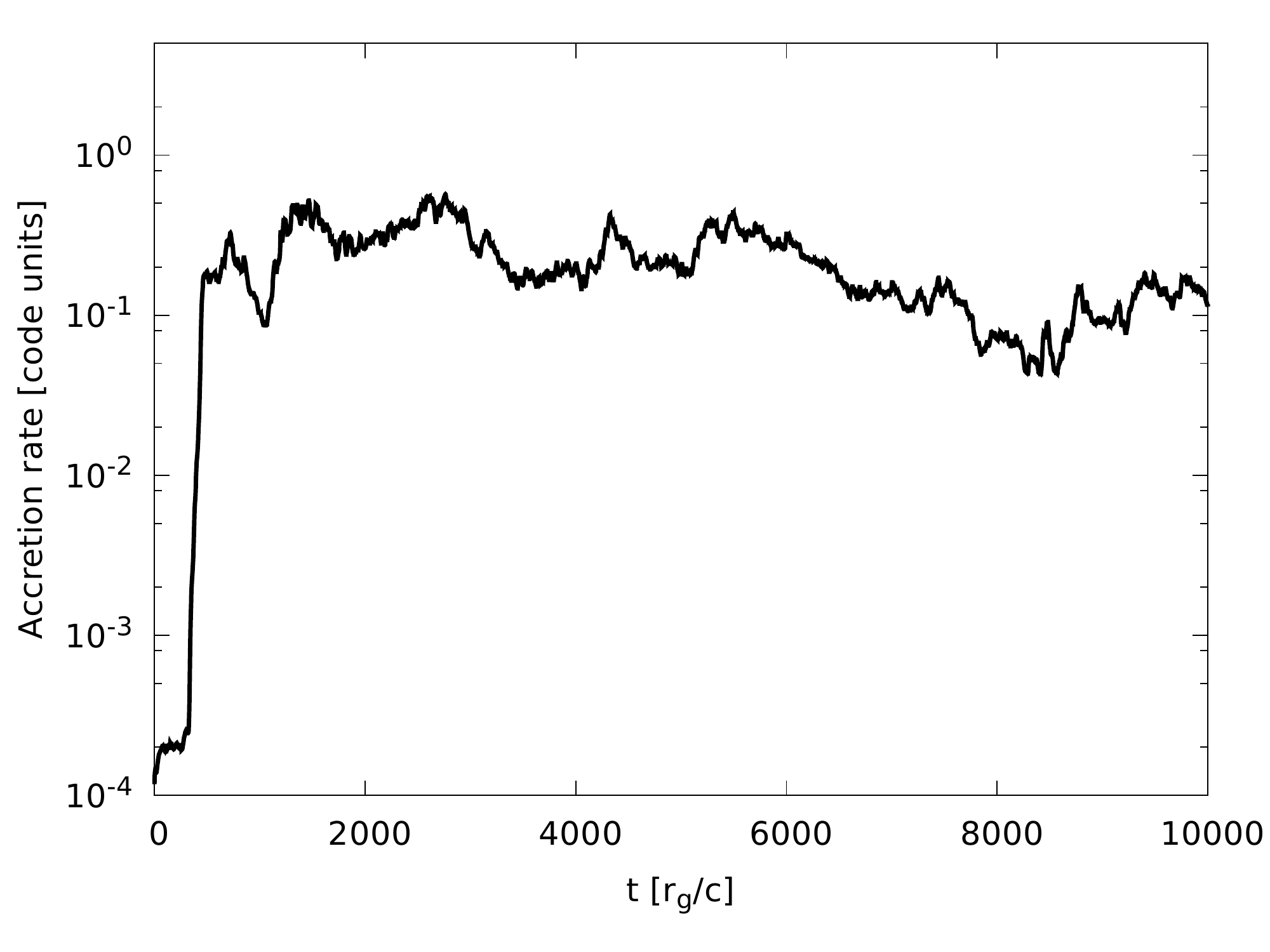}
    \includegraphics[width=0.49\textwidth]{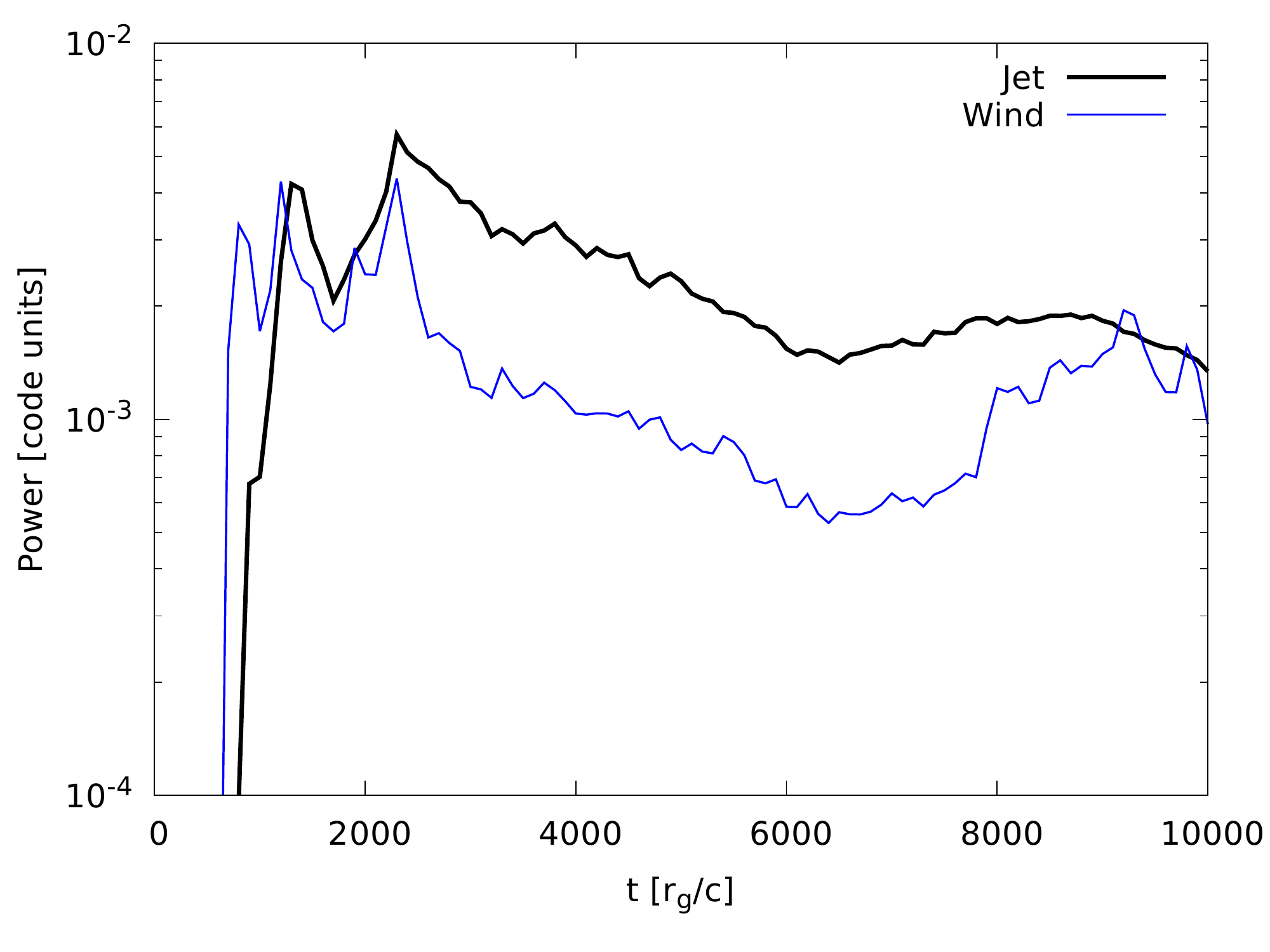}
    \caption{Left panel: Accretion rate in code units as a function
      of time. Right panel: Jet and wind power in code units as a
      function of time.}
    \label{fig:acc-ljet}
    \includegraphics[width=0.95\textwidth]{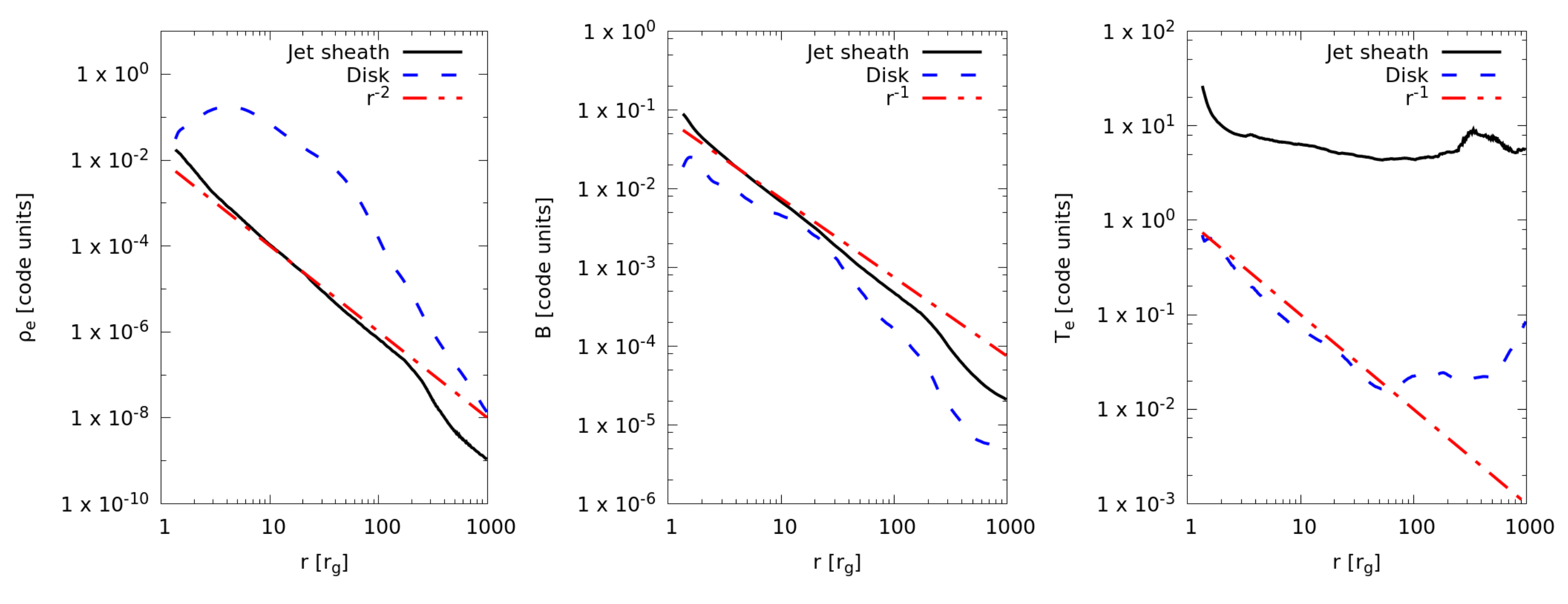}
    \caption{Radial profiles of the dimensionless electron density
      (left), magnetic-field strength (middle) and electron temperature
      (right). Black lines correspond to jet averaged quantities, dashed
      yellow lines to disk-averaged quantities and the red dashed lines
      correspond to power-law profiles predicted in analytical works by
      \cite{blandford1979} and \cite{falcke1995}. Also, the jet sheath is close to isothermality.}
    \label{fig:lineprof}
\end{figure*} 

A representative snapshot of the simulation is shown in
Fig. \ref{fig:snapshots}. The simulation produces a well-resolved
relativistic jet up to the edge of the simulation domain at $1000 ~\rg$
in the $z$-direction. At $z=40~\rg$ the jet diameter is resolved with 160
cells, and with 32 cells at $z=1000~\rg$. The accretion rate through the
event horizon is shown in the left panel of Fig.~\ref{fig:acc-ljet} as a
function of time; note that the accretion rate initially increases
sharply, and then settles around $0.2$ (in code units) at later
times. The jet and wind power are instead shown in the right panel of
Fig.~\ref{fig:acc-ljet}; both of the quantities are calculated by
performing the integral over the constant $r=100~\rg$ surface
\begin{equation}
\dot{E} = \int_0^{2\pi}\int_{0}^{\pi} (- T^r_t - \rho u^r)\,
\chi_{(\cdot)}\, \sqrt{-g} \, d\theta \, d\phi
\end{equation}
where the function $\chi_{(\cdot)}$ selects only material in the jet,
wind, or disk following the setting 
\begin{subequations}
   \begin{align}
    \chi_{\rm jet} &= (b^2/\rho >1 ~{\rm or}~ \mu > 2)\\
    \chi_{\rm wind} &= ({\rm not} ~\chi_{\rm jet} ~{\rm and}~ -h u_t > 1) \\
    \chi_{\rm disk} &= ({\rm not} ~\chi_{\rm jet} ~{\rm and} ~{\rm not}~ \chi_{\rm wind}),
\end{align}
\end{subequations}
and $\mu$ denotes the energy flux normalized to the rest-mass energy in 
the radial direction  $\mu=(-T^r_t-\rho u^r)/(\rho u^r)$. Hence, 
the jet is defined as the region which reaches
asymptotic Lorentz factors of at least 2. The optional condition
$b^2/\rho>1$ also selects the flow in the inner axial region, where the
Poynting flux necessarily vanishes. The disk wind is then the remaining
unbound material and the disk itself is composed of the bound material.

Analytic work on radial profiles of relativistic jets was performed by
\cite{blandford1979} and subsequently by \cite{falcke1995}. In these
Blandford-K\"onigl jet-models, the electron density decreases as a
function of radius as $\rho_{\rm e}\propto r^{-2}$, the magnetic field
strength as $B \propto r^{-1}$, and the equipartition electron
temperature in the jet is constant. The temperature in the disk is set by
the virial theorem, and follows $T_e \propto r^{-1}$. To compare our
simulations with these analytical formulae, we compute averages on
spherical shells at different fixed radii of the electron density $\rho_e
$, magnetic-field strength $B$, and electron temperature $T_e$. This is
done by performing the following integral
\begin{equation}
  q(r)  = \frac{1}{\Delta t} \int \left(\frac{\int
  \int_0^{2\pi} q(t,r,\theta,\phi) \sqrt{-\gamma(r,\theta)} d\theta d\phi}{\int
  \int_0^{2\pi} \sqrt{-g(r,\theta)} d\theta d\phi} \right) dt.
\end{equation}
where $\gamma(r,\theta)$ is the determinant of the three metric
 The integral in the $\theta$-direction depends on the
local plasma criteria. We consider two regions of interest; a jet sheath,
for which $0.1 < \sigma < 5.0$, and the accretion disk, $\sigma < 0.For$.
The time average, on the other hand, is performed using snapshots of the
simulation between $t=5000 ~\tunit$ and $t=10^4 ~\tunit$, with a total of
hundred snapshots. The computed radial profiles are shown in
Fig.~\ref{fig:lineprof} and are over-plotted with the analytic
predictions \citep{blandford1979,falcke1995}. The
equipartition electron temperature in the jet (right panel) shows a flat
profile up to $200~\rg$, followed by is an increase of temperature that
correlates with the break in the profile of the electron density. The
break is caused by de-collimation of the jet, whose origin could be due
to the limited initial size of the torus. Note that the wind emitted by
the disk effectively acts as a collimation agent; however, because of its
limited size, the collimation stalls at radii $r>200 ~\rg$.

\subsection{Spectra and synchrotron images: dependency on electron distribution function}

In this Section, we discuss the spectral energy distributions (SEDs) of
our thermal-jet and $\kappa$-jet models. The SEDs are calculated at an
inclination of $i=160^\degree$, which ensures that the emitting region is
in the South, as suggested by the EHT results
\citep{eht-paperI,eht-paperV}. Furthermore, the field-of-view of the
camera is set to be $1000~\rg$ in both the $x$ and $y$-directions, while
the resolution is set to be $2000 \times 2000$ pixels.

\subsubsection{Fitting the SED}

\begin{figure*}
  \centering
  \includegraphics[width=0.97\textwidth]{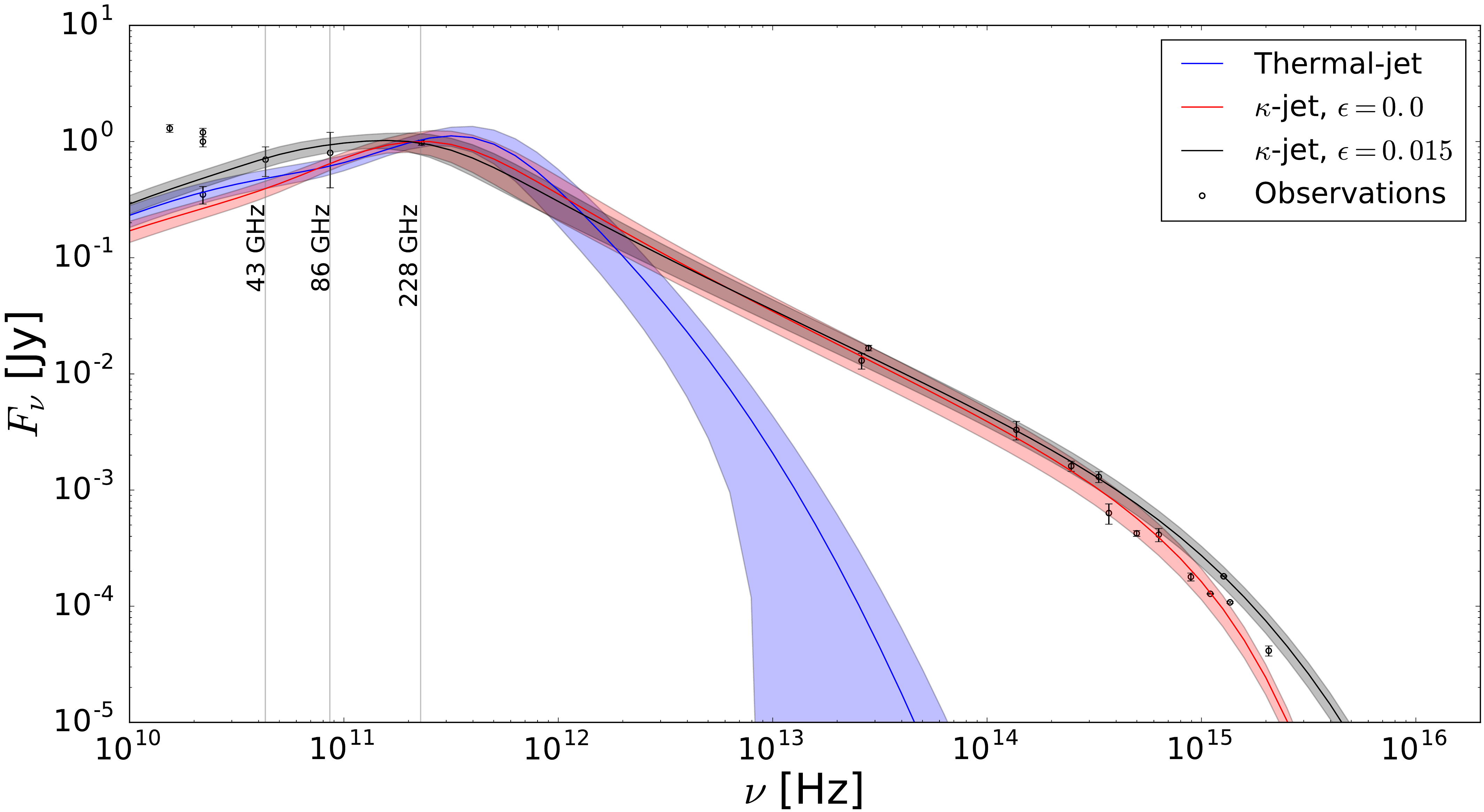}
       \caption{SEDs for the thermal-jet (orange) and $\kappa$-jet
         (black) with their corresponding rms, overplotted with
         observational data points by
         \cite{doeleman2012,akiyama2015,prieto2016,kim2018}.}
        \label{fig:spectrum}
\end{figure*}

After averaging in time the SEDs from our models between $t=5000 ~\tunit$
and $t=10^4 ~\tunit$), these have been fitted to non-simultaneous
observations by
\cite{doeleman2012,akiyama2015,prieto2016,walker2018,kim2018}. The fit
parameters are shown in Table \ref{tab:modelpar}, which highlights that
the thermal-jet and $\kappa$-jet models differ in the accretion rate by a
factor $\approx 2$. The corresponding SEDs are shown in
Fig.~\ref{fig:spectrum}, which shows that $\kappa$-jet models recovers
well the NIR flux. In particular, when comparing the $\epsilon=0.0$ and
the $\epsilon=0.015$ models (the latter uses an injection radius of
$r_{\rm inj}=10~\rg$ and has a slightly lower accretion rate), it is
possible to appreciate that the $\epsilon=0.015$ model has a larger and
flatter radio spectrum at frequencies below $\nu=228$ GHz.

\begin{table}
  \label{tab:modelpar}
\centering
\begin{tabular}{cccc} 
 \hline
 \hline
 Parameter  & Thermal& $\kappa$, {$\epsilon=0$}  & $\kappa$ , {$\epsilon=0.015$}\\
 \hline
$i$ & $160^{\degree}$ & $160^{\degree}$ & $160^{\degree}$\\ 
$\mathcal{M}~[{\rm g}]$  & $1.8\times10^{29}$ & $10^{29}$ & $8\times10^{28}$\\
$P_{\rm jet} \,{\rm ~[erg~s^{-1}]}$ & $1.1\times10^{43}$ & $5.9\times10^{42}$  & $4.7\times10^{42}$ \\
$\langle \dot{M} \rangle_t \,~[\mdotu]$ & $8.4 \times 10^{-3} $ & $4.7 \times 10^{-3} $ & $3.8 \times 10^{-3} $ \\
$B_0 ~[G]$ &  $1.6\times10^3$ &  $1.2\times10^3$ & $1.1\times10^3$ \\
$n_0 ~[{\rm cm}^{-3}]$ &  $1.34\times10^8$ &  $7.5\times10^7$  & $6\times10^7$ \\
$R_{\rm high}$ &  100 &   100  & 100\\
$R_{\rm low}$ &   1  & 1 & 1\\
%$\epsilon$   & - & $0$ & $0.015$ \\
$r_{\rm inj}$ &  - &  - & $10~\rg$\\
 \hline
\end{tabular}
\caption{List of parameters are used in the radiative-transfer
  simulations. }
\end{table}

After $228$ GHz both $\kappa$-jet models recover a power-law with an
index of $\alpha \approx -0.7$, where $\alpha=-(p-1)/2$ for a power law
distribution of non-thermal electrons $F_\nu \propto
\nu^\alpha$. Furthermore, when compared to the thermal model, the flux in
the $\kappa$-jet models is higher at lower frequencies ($\nu<10^{11}$ Hz)
and at the higher frequencies ($5\times10^{12} {\rm Hz} <\nu$).

When considering the various cut-off models, the cooling cut-off turned
out to be unimportant, in agreement with the findings of
\citep{moscibrodzka2016,broderick2015}. On the other hand, the for the
synchrotron burnoff, the correct cut-off is obtained if $E/B\approx
10^{-6}$, but no physical model is possible that recovers such a
ratio. The only criterion that recovers the cut-off frequency is the
Hillas criterion, which is obtained when the plasmoid size is set to
$L\approx 10^5 - 10^7$ cm, depending on the local magnetic field
strength.

\subsubsection{Synchrotron maps}
\begin{figure*}
%\centering
    \includegraphics[width=0.95\textwidth]{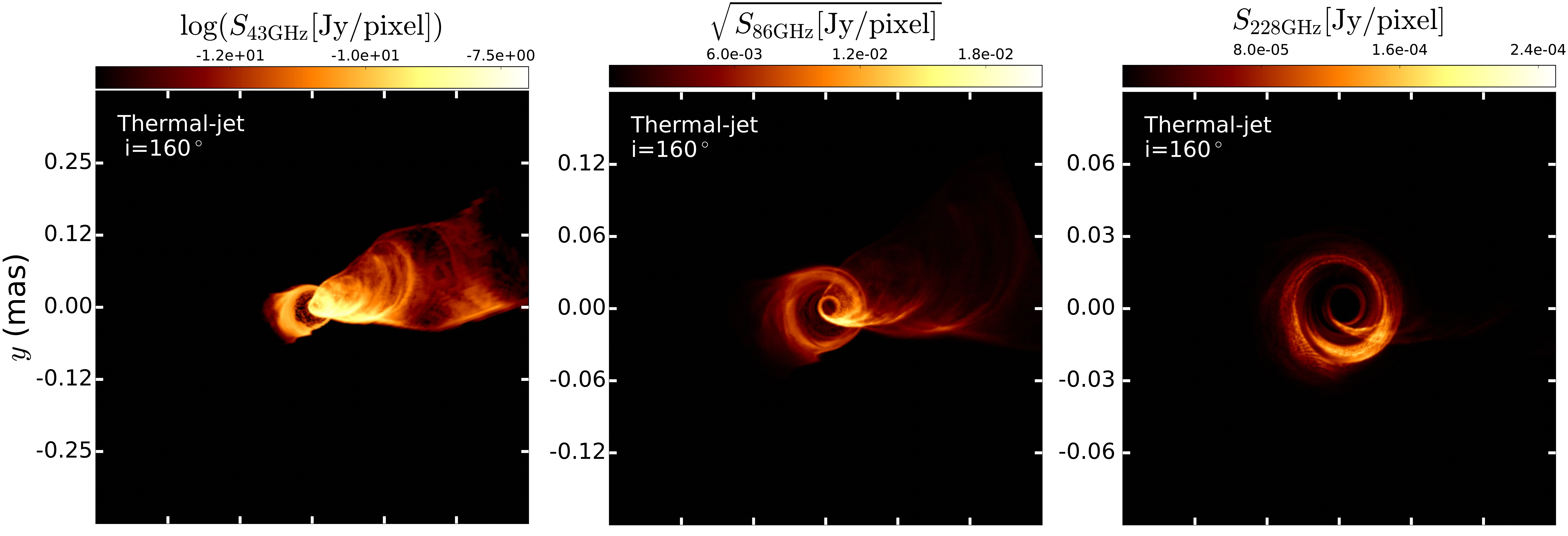}
    \includegraphics[width=0.95\textwidth]{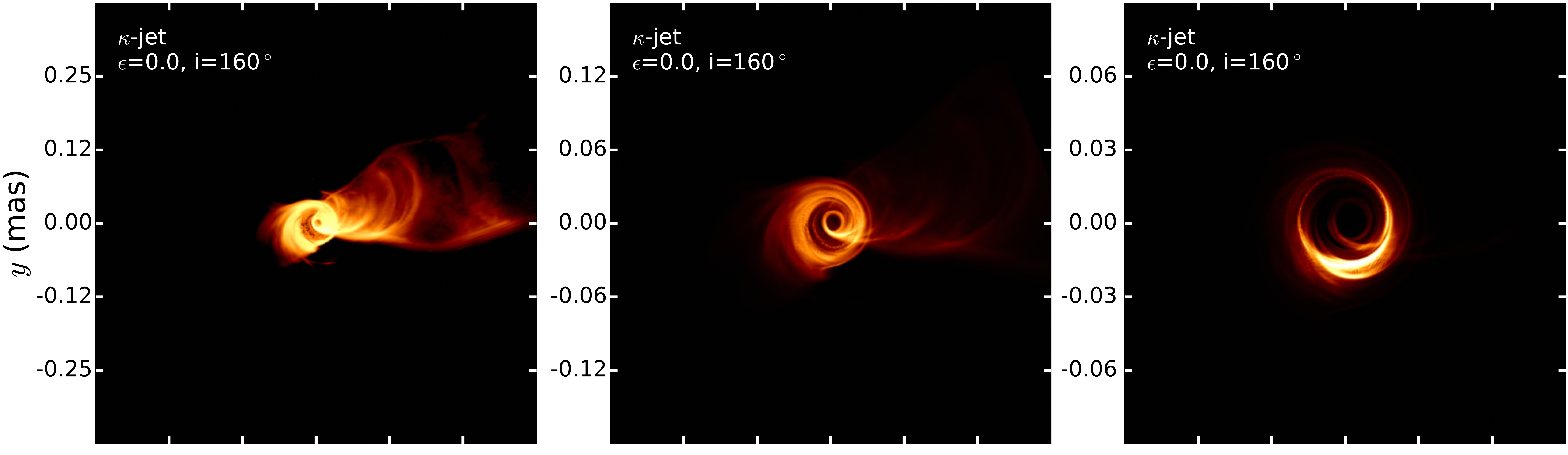}
    \includegraphics[width=0.95\textwidth]{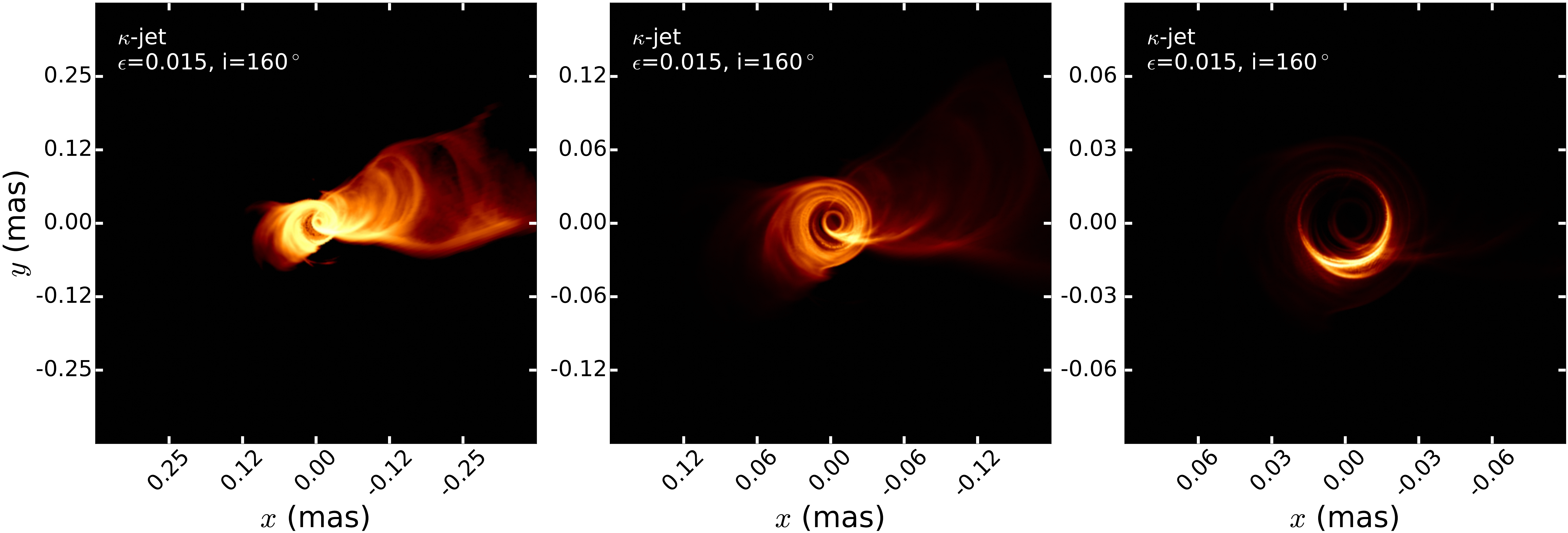}
    \caption{ {From left to right: $43, 86$, and $228$ GHz}. Top row:
      synthetic images at a single snapshot of the thermal-jet at an
      inclination of $i=160^\degree$. Second row: same as top row but for
      the $\epsilon=0.0$ $\kappa$-jet. Bottom row: same as the first and
      second row but for the $\epsilon=0.015$
      $\kappa$-jet.}  \label{fig:img-160}
    \includegraphics[width=0.95\textwidth]{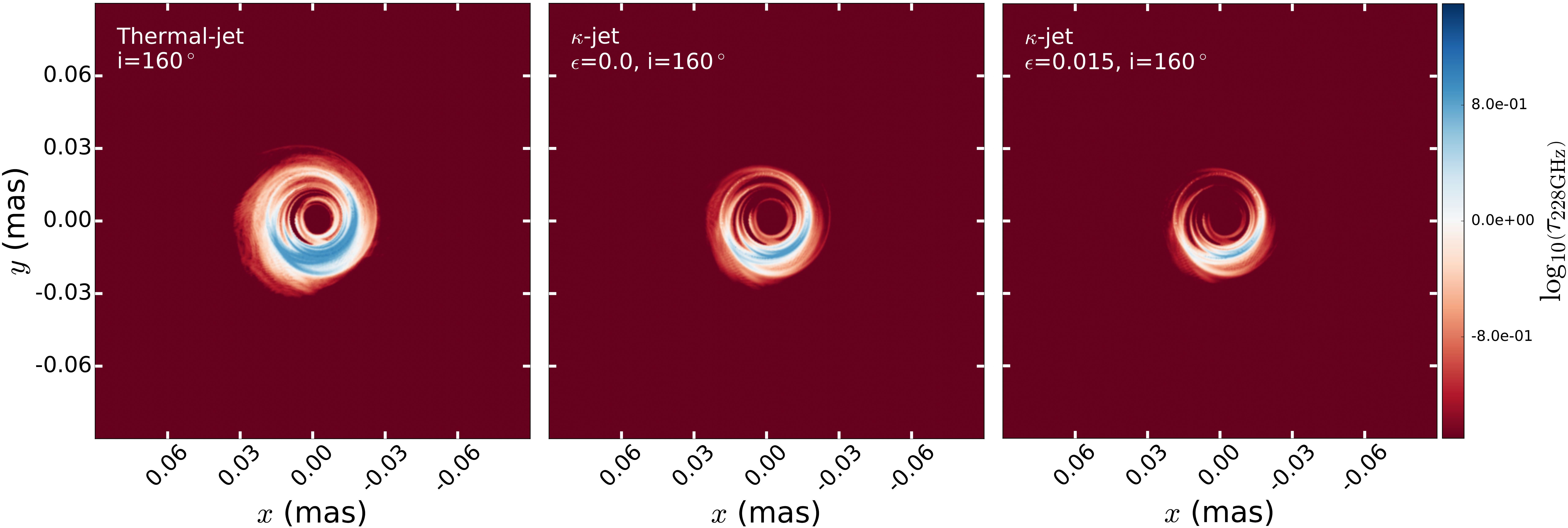}
    \caption{Logarithmic optical-depth maps at {228 GHz} of a single snapshot of the models at an inclination of $i=160^\degree$.}
                  \label{fig:tau-160}
\end{figure*}

The synthetic synchrotron maps are computed at three frequencies: $43,
86$, and $228$ GHz. The same inclination used for the SEDs is employed
here and the images for the thermal case are shown in the top rows of
Fig. \ref{fig:img-160}, with the the $\kappa$-jet models shown in the
second and third rows. The maps shown are computed with a single GRMHD
snapshot at $t=10^4~\tunit$. The forward jet at 43 GHz is aligned with
the observed jet position angle at 43 GHz VLBI observations
\citep{Janssen2019}, namely, $250^\degree$. The assumed mass and distance
are $M_{\rm BH} = 6.2 \times 10^9 ~\msun$ \citep{gebhardt2011} and $d =
16.7$ Mpc \citep{mei2007}, which results in a field of view of: $0.744,
0.372$ and $0.186$ mas for the $43, 86$, and $228$ GHz maps, respectively.

The thermal-jet and $\epsilon=0.0$
$\kappa$-jet model show a similar source morphology at $43$ GHz and $86$ GHz,
and $\epsilon=0.015$ $\kappa$-jet model is more extended in jet
length. At $228$ GHz both $\kappa$-jet models deviate from the
thermal-jet model, the width of the ring around the shadow decreases when
particle acceleration is present. {In all 228 GHz images two rings are
  visible, the outer ring is the photon ring and marks the shadow of the
  black hole, the fainter smaller ring is emission originating from the
  jet facing the observer, see Appendix \ref{appendixA} for more
  details.}

%% \subsubsection{Optical depth maps}

The logarithmic optical-depth maps at $228$ GHz are shown in Fig. \ref{fig:tau-160},
where the size of the optically thick region (in blue) decreases when
particle acceleration is present. This is in agreement with the less
extended structure visible in the intensity-maps of
Fig. ~\ref{fig:img-160}. The reason behind this behaviour is that lower
mass-accretion rates decrease both the density and the magnetic field
strength, hence decreasing the optical thickness of the jet base. As a
result, for any given frequency, accelerated particles at lower
mass-accretion rates contribute more than their thermal counterpart. 

\subsubsection{Origin of the jet emission}

To obtain a quantitative understanding of how much flux originates either
from the forward or the counter-jet, i.e., the jet facing away from the
observer, we computed synthetic images where the emission coefficient was
set to zero either in the southern or northern hemisphere, while keeping
the absorption coefficients in place. We computed the time-averaged ratios and the spread of the
flux originating from the southern to flux from the northern hemisphere of
both our models for all slices between 5000 and $10^4~\tunit$
at $43, 86$, and $228$ GHz and have reported them in Table
\ref{tab:ratios}. When electron acceleration is present,
the overall trend is that at $43$ and $86$ GHz, the ratio shifts to the
counter-jet, while at $228$ GHz no large shifts are seen. We therefore conclude that  the
counter jet at  $43$ and $86$ GHz is more dominant in
the $\kappa$-jet models compared to the thermal models.
Appendix \ref{appendixA} provides a simple phenomenological model
that is capable of reproducing this effect, where it is caused 
by a combination of gravitational lensing and the blocking of light 
by the black-hole's event horizon.

\begin{table}[h!]
\centering
\begin{tabular}{lccc} 
 \hline
 \hline
 & 43 GHz & 86 GHz & 228 GHz \\
\hline
Thermal-jet  & $8.3\pm2.2$ & $2.3\pm0.6$  & $0.4\pm0.1$  \\
$\kappa$-jet, $\epsilon=0.0$ & $2.6 \pm 0.7$ & $0.9 \pm 0.2$ & $0.3 \pm 0.09$  \\
$\kappa$-jet,  $\epsilon=0.015$ & $2.6 \pm 0.7$ & $1.1\pm0.3$ & $0.5 \pm 0.2$ \\
\hline
\end{tabular}
\caption{Table showing the ratio in flux between the forward and
  counter jet at $43, 86$, and $228$ GHz for the thermal-jet and
  $\kappa$-jet models.}
\label{tab:ratios}
\end{table}
%

%% For our images computed at 228 GHz it is observed that most of the
%% emission that reaches the observer originates from the `counter-jet',
%% i.e., the jet facing away from the observer. For a simple phenomenological
%% model that is capable of reproducing this effect, which is caused by a
%% combination of gravitational lensing and the blocking of light by the
%% black-hole's event horizon, see 

\subsubsection{Core size and shift}

We computed the source size of our models at $43, 86$, and $228$ GHz by
using image moments \citep{hu1962}. The sources sizes are computed over a
range of $5000$ to $10^4 ~\tunit$ and in Table \ref{tab:sizes} we report
the time-averaged major and minor full-width half maxima (FWHM) and their
corresponding spread.

\begin{table}[h!]
%  \label{tab:major}
\centering
\begin{tabular}{lccc} 
 \hline
 \hline 
$\theta_{\rm major}$ ($\mu$as) & ${43 ~{\rm GHz}}$  & ${86 ~{\rm GHz}}$  & ${228 ~{\rm GHz}}$  \\
\hline
Thermal-jet  & $141\pm25$ & $87\pm 12$  & $45\pm4$  \\
$\kappa$-jet, $\epsilon=0.0$ & $128\pm20$ & $73\pm9$ & $41\pm3$  \\
$\kappa$-jet, $\epsilon=0.015$ & $142\pm18$ & $87\pm10$ & $53\pm6$ \\
\hline
\hline 
$\theta_{\rm minor}\ (\mu$as) & & &  \\
\hline
Thermal-jet  & $56\pm 4$ & $43\pm2$  & $33\pm1$  \\
$\kappa$-jet, $\epsilon=0.0$ & $54\pm3$ & $43\pm1$ & $32\pm1$ \\
$\kappa$-jet, $\epsilon=0.015$ & $60\pm3$ & $48\pm2$ & $36\pm2$  \\
\hline
\end{tabular}
\caption{Top: FWHM along the major axis for the thermal-jet and
  $\kappa$-jet at $43, 86$, and $228$ GHz. Bottom: same as top
  but along the minor axis.}
\label{tab:sizes}
\end{table}

\begin{figure}[h!]
\centering
    \includegraphics[width=0.45\textwidth]{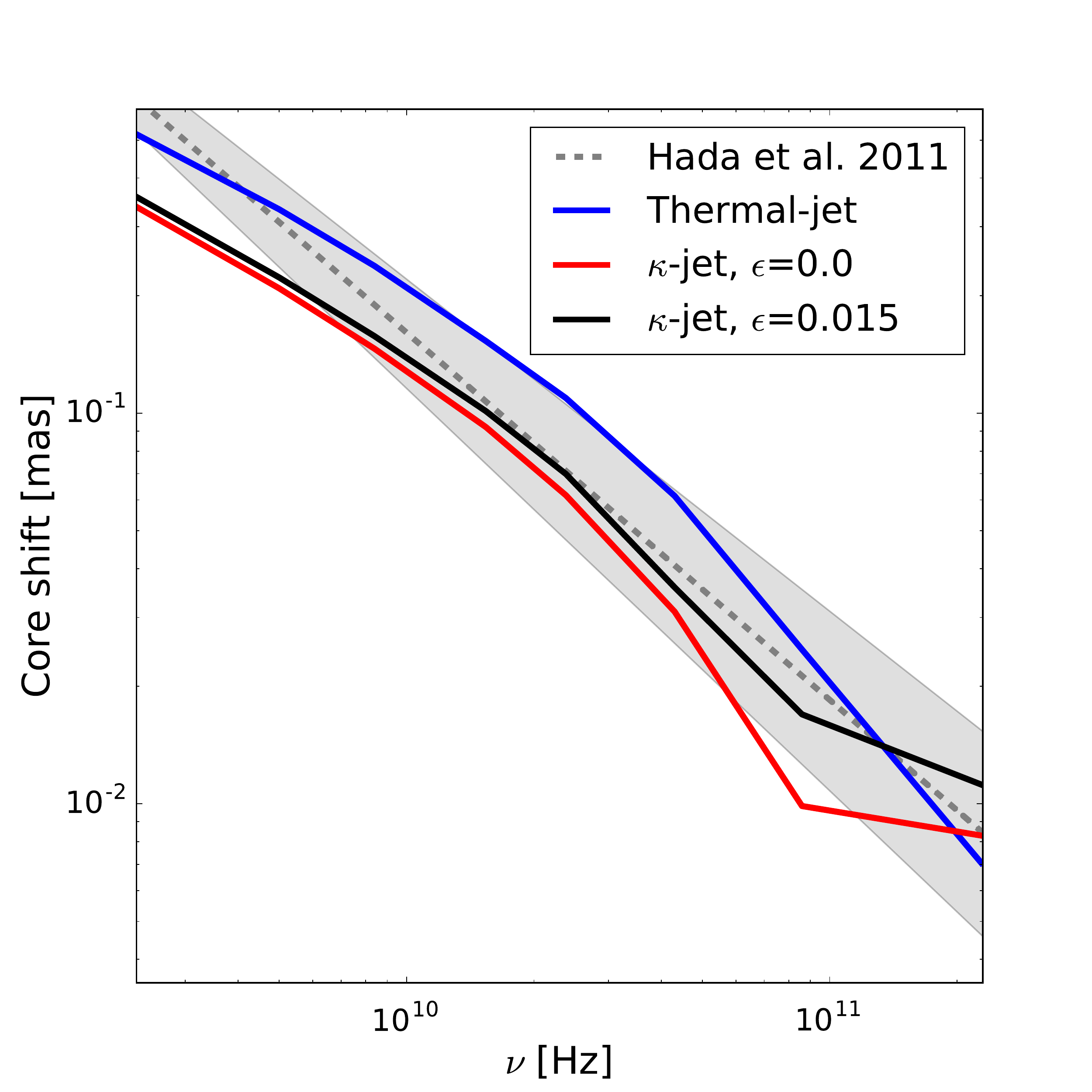}
       \caption{ RA offset from the 43 GHz core as a function of
         frequency. Orange triangles correspond to a thermal-jet, black
         dots to a $\kappa$-jet, grey line represent the observational
         fit $r_{\rm RA}(\nu) = (1.4\pm0.16) \nu^{-0.94\pm0.09}$ to the
         M\,87 core by \cite{hada2011}.}
        \label{fig:coreshift}
\end{figure}

We computed the core shift with respect to the black-hole's center at the
following observational frequencies; $2.3, 5, 8.4, 15.4, 23.8, 43, 86$,
and $228$ GHz. The core shift was calculated by computing the first-order
image moments of time-averaged images and the comparison of the values
obtained with the observational fit of \cite{hada2011}, i.e., $r_{\rm
  RA}(\nu) = (1.4\pm0.16) \nu^{-0.94\pm0.09}$, is shown in
Fig.~\ref{fig:coreshift}. The observed core shift is in agreement with
the analytical predictions for which the core position should shift for a
conical jet as a function of frequency as $r_{\rm core} \propto \nu^{-1}$
\citep{blandford1979,falcke1995,davelaar2018}, and in agreement with
simulations of collimated jets \citep{porth2011}. The $\kappa$-jet models
show smaller core shifts with respect to the thermal-jet model, probably
because the counter-jet is more dominant.

\subsubsection{Comparison with 43 GHz data}

Finally, we compared our thermal-jet and $\kappa$-jet models with the 43
GHz VLBI observations, where M\,87 was tracked for 8 hours with all VLBA
stations \footnote{PI: R. Craig Walker, project code: BW0106}. The 
data was recorded with a bandwidth of 256\,MHz, with
the calibration and imaging of the data having been described by
\citet{Janssen2019}.

\begin{figure*}[h!]
    \includegraphics[width=0.95\textwidth]{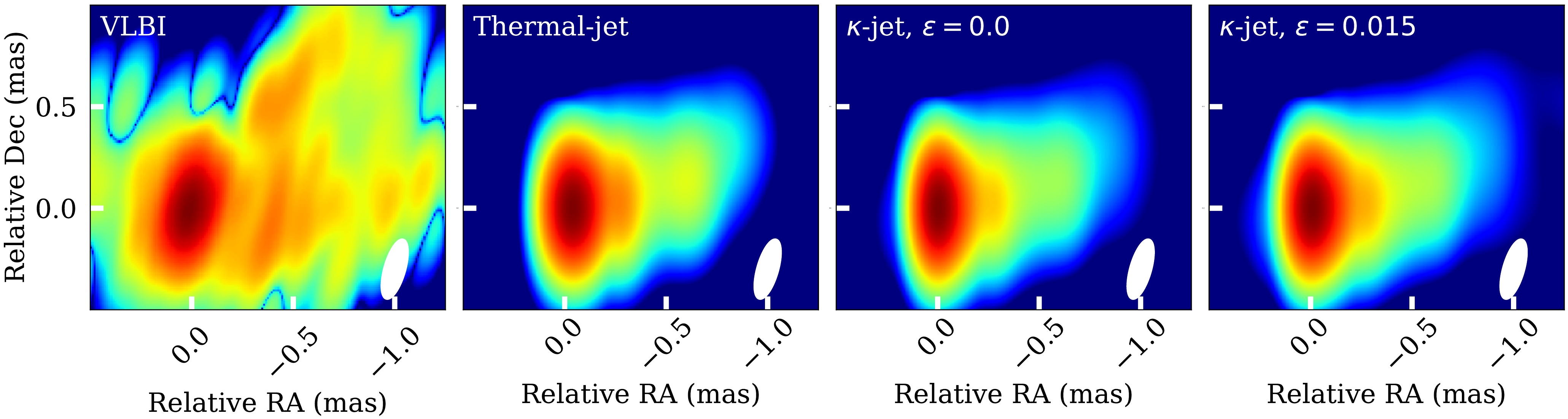}
          \caption{First panel from the left: 43 GHz radio map of M\,87
            \citep{Janssen2019}. Second panel: synchrotron map of the
            thermal-jet model , convolved with a 2D Gaussian beam. Third
            panel: same as the second panel but now for a $\kappa$-jet
            model with $\epsilon=0.0$. Fourth panel: Same as the third
            panel but now with $\epsilon=0.015$. The white ellipse
            indicates the beam used to convolve the images. All models
            produce a jet that is too narrow compared to the VLBI
            map. The extent of the jet increases when electron
            acceleration is present, and is maximum for
            $\epsilon=0.015$.}
        \label{fig:img-obs-data}
\end{figure*}

To compare with this observational data, we re-computed synthetic images
with a large field of view of 3.7 mas and convolved them with a
$0.3\times 0.1\,{\rm mas}^2$ beamsize by using the eht-imaging library
\citep{chael2016b,chael2018b}. The result of the comparison can be seen
in Fig.~\ref{fig:img-obs-data} and highlights that the $\kappa$-jet
models show more extended structure with respect to the thermal-jet
model. Note that at 43 GHz all models deviate from the VLBI observations
at larger scales. Furthermore, in the observed image the flux levels
upstream of the jet are higher and the jet opening angle is wider.

\section{Discussion}
\label{sec:disc}

\subsection{CKS GRMHD simulations}

Current models of the radio emission near the supermassive black hole in
M\,87 are based on GRMHD simulations using spherical polar
coordinates. In this work, we used instead Cartesian coordinates, which
do not require specialized treatment of the polar axis which represents a
coordinate singularity (of the inverse metric) in spherical
coordinates. The addition of AMR resulted in a highly resolved jet
region, the jet diameter at $z=40 ~\rg$ is resolved by 160 cells and at
$z=1000 ~\rg$ is resolved by 32 cells. The obtained jet resolution is
well above the values of 20-26 cells per jet radius reported in
convergence studies of jets by \cite{anjiri2014,massaglia2016}. We
computed the mass-accretion rate and radial profiles of density,
magnetic-field strength and temperature which are consistent with their
spherical counterparts, the comparison of this can be found in
\cite{PorthChatterjeeEtAl2019}. The downside of the Cartesian grid is
that in spherical grids it is possible to use a logarithmic grid in the
radial direction, which results in higher resolutions close to the
horizon. To ensure that the MRI is not under-resolved, we have computed
the relativistic MRI quality factors \citep{noble2010} finding they are
normally above ten in the bulk of the accretion disk, thus satisfying the
requirements for a sufficiently resolved MRI found by
\cite{Sano2004}. Furthermore, in parallel works
\citep{Olivares2019,PorthChatterjeeEtAl2019} it was shown that the
Cartesian CKS simulations show a behaviour in the nonlinear phase that is
similar to the spherical simulations.

\subsection{The effect of electron acceleration on the SED}

We computed spectra for our thermal-jet model and $\kappa$-jet models,
wherein the latter case we parametrize the power-law index of the eDF
based on sub-grid PIC simulations by \cite{ball2018}. The addition of
accelerated electrons in the jet sheath leads to a very good fit to the
observational data from radio up to the NIR. Our $\kappa$-jet model is an
extension of the model presented by \cite{moscibrodzka2016}, which only
studied the thermal-jet case. Our models use their best-fit inclination
angle of $i=160^\degree$, such that the emitting region is in the South
and the orientation of the asymmetry is in agreement with the image in
\cite{eht-paperI}. The radio SED shows a flat spectrum in both the
thermal and $\kappa$-jet models. This is consistent with more recent work
by \cite{ryan2018,chael2019}, who have evolved the thermal electron
population as a separate fluid in the GRMHD simulation.

In contrast to previous works, our $\kappa$-jet models also recover the
observed NIR flux by extending the optically thin emission with a
power-law. The results are similar to the ones presented by
\cite{dexter2012}, who also injected accelerated electrons based on the
amount of available magnetic energy. The $\kappa$-jet models yield a jet
power of the order of $10^{43}$ ergs s$^{-1}$, which is in agreement with
observations of the jet core power by \cite{reynolds1996}, and is
approximately two times lower than the thermal-jet models. This is
probably due to the fact that in the $\kappa$-jet models there is a
larger contribution of electrons in the tail of the distribution
functions with respect to the thermal-models. Since these electrons emit
at higher $\gamma$ values, this results in a higher flux contribution per
unit mass.

After defining the radiative efficiency as $\epsilon_{\rm
  rad}=L/\dot{M}c^2$, we found that the thermal-jet has $\epsilon_{\rm
  rad}=0.003$. This is to be contrasted with $\epsilon_{\rm rad}=0.013$
and $\epsilon_{\rm rad}=0.020$ for the $\kappa$-jet models with
$\epsilon=0.0$ and $\epsilon=0.015$, respectively. An important note 
is that we do no include X-ray emission in this work. Although, 
the obtained values are well below the thin disk efficiency, thus justifying our assumption that the radiation can be decoupled from the evolution of the dynamics of the plasma.

\subsection{The effect of electron acceleration on synchrotron maps}

At 43 and 86 GHz, both $\kappa$-jet models show a more dominant
counter-jet when compared to the thermal-jet model, hinting to a
behaviour that could be observable by future GMVA-ALMA
observations. There is also a clear difference in the extent of the
emission of the forward jet in the $\epsilon=0.015$ $\kappa$-jet model
when compared to the $\epsilon=0.0$ $\kappa$-jet and to the thermal-jet
model, with the emitting region being more compact in the $\kappa$-jet
models at $228$ GHz. The reason for this is that there is more energy
available at higher $\gamma$ in the eDF, which results in a higher flux
contribution per unit mass. Indeed, to obtain a fit to the data, a lower
mass-accretion rate is needed. Since our mass-accretion rate sets the
scaling of the densities and magnetic fields, it also changes the optical
thickness of the source. As a result, a more optically thin model will
show a more compact emission region.

A comparison with the result from \cite{moscibrodzka2016} shows that
similar source morphologies at all frequencies for the thermal
model. However, at $228$ GHz our images show a more optically thick inner
ring feature that partially blocks the view to the shadow. The reason for
this is that our initial conditions differ from those of
\cite{moscibrodzka2016}, as they used a disk with a pressure maximum at
$24 ~\rg$, resulting in an outer radius of $r=240 ~\rg$, while we used a
pressure maximum at $12 ~\rg$ and outer radius of $r=40 ~\rg$. A larger
disk is initially seeded with larger toroidal magnetic-field loops, and a
larger loop increases the magnetic flux at the horizon at later
times. These stronger magnetic fields will affect the overall source
morphology, resulting in wider opening angles which lead to less
obscuration of the shadow by the forward jet.

\subsection{Core size, shift, and jet opening angle}

The obtained core sizes for our models are close to the observed values:
$ \theta^{43 ~{\rm GHz}} = 0.13 \pm 0.01$, $ \theta^{86 ~{\rm GHz}} =
0.079 \pm 0.021$ \citep{hada2013}, and $ \theta^{228 ~{\rm GHz}} = 0.040
\pm 0.002$ \citep{doeleman2012}. If we compare these to values reported
in Tables \ref{tab:sizes}, we find that our models at 43 and 86 GHz are
within the error margins of the observations. At 228 GHz, the
$\epsilon=0.0$ $\kappa$-jet {recovers the observational value}. The
thermal-jet model is slightly larger, this is probably caused by the
larger emission region around the shadow. In the $\epsilon=0.015$ case,
the deviation is caused by a more pronounced jet feature.

We obtain core-shift relations for both our models by calculating the
core position that follows the trend found by \cite{hada2011}. They
computed the core shift with respect to the $43$ GHz core. Their obtained
fit is then extrapolated to higher frequencies where they find an offset
of 40 $\mu$as at 228 GHz. At frequencies below $10$ GHz, deviations with
the fit from \cite{hada2011} are present. A possible explanation for this
is the limited simulation domain of $1000~\rg$ and the de-collimation of
the jet after $r\approx300~\rg$.

An important remark to make is that we have here considered a Standard
And Normal Evolution (SANE) simulation. This results in a low magnetic
flux at the event horizon when compared to Magnetically Arrested Disc
(MAD) simulations that result in the maximum amount of flux that can
penetrate the event horizon \citep{narayan2003,tchekhovskoy2011}. If we
compare our results with the MAD simulation from \cite{chael2019}, our
jet opening angle is smaller and our models are inconsistent with the
observational constraints on the jet opening angle at 43 GHz
($55^\degree$ \citealt{walker2018,Janssen2019}); by contrast,
\cite{chael2019} showed that their thermal MAD simulations do match the
observed opening angle.

%-->

\subsection{Reconnection as the source of particle acceleration}

The electrons' energy-distribution function is one of the key open
questions in modeling the appearance of jets launched by supermassive
black holes. Simulations of these acceleration mechanisms rely on
non-ideal effects, which are not captured in GRMHD-based
simulations. Fully resistive treatments of the plasma using non-ideal
GRMHD simulations
(\citealt{Palenzuela:2008sf,Ohsuga2009,Dionysopoulou2013,Bucciantini2012,delzanna2016,qian2017,qian2018,ripperda2019})
or general-relativistic PIC simulations
(\citealt{watson2010,levinson2018,parfrey2018}) are being developed and
will help to provide detailed answers to these questions in the
future. In principle, alternative acceleration mechanisms could be at
work, such as shocks. In our model, the main region of emission is where
the magnetization $\sigma$ is around unity, where shocks are known to be
less efficient \citep[see e.g.,][]{sironi2015}.

\subsection{The Event Horizon Telescope results}

In \cite{eht-paperV}, GRMHD models were used to interpret the first image
of a black hole. In the post-processing of the GMRHD models, only a
thermal distribution function were considered. In this work, we show the
effect of electron acceleration by performing a comparison with a purely
thermal model. The overall trend is that the emission region is optically
thinner and smaller in size. Also the accretion rates and jet-power drop,
which could have implication for some of the models reported in
\cite{eht-paperV}. A detailed comparison with respect to the EHT data is
beyond the scope of this work, but realistic synthetic data generation
based on the models presented here will be discussed by
\cite{Roelofs2019}.

\section{Conclusion}\label{sec:conc}
We have presented a $\kappa$-jet model for the accreting black hole in
M\,87 based on an AMR GRMHD simulation in Cartesian-Kerr-Schild
coordinates, coupled to radiative-transfer calculations that include
sub-grid models for electron acceleration based on reconnection in the
magnetized jet. {The use of a Cartesian grid with AMR resulted in a
  high-resolution jet simulation that we used to model the jet launching
  point in M\,87.} We have demonstrated that we can obtain a fit to the
M\,87 SED from radio up to NIR if there is an accelerated electron
population present in the jet. The model does not evolve the electron
distribution function in time and does not include cooling; both of these
aspects will be considered in future works. The jet opening angle at
43GHz is too narrow, \cite{chael2019} showed that a MAD type accretion
disk can recover this opening angle, and we plan to explore this setup in
future works with the addition of particle acceleration. The model
reproduces the broadband SED from radio up to NIR, observed source sizes,
core shifts and recovers a jet power that is consistent with
observations.

\section{Acknowledgments}

The authors thank M. Moscibrodzka, C. Gammie, A. Philippov, Z. Younsi,
and B. Ripperda for valuable discussions and feedback during the
project. This work was funded by the ERC Synergy Grant
``BlackHoleCam-Imaging the Event Horizon of Black Holes'' (Grant 610058,
\cite{goddi2017}). The GRMHD simulations were performed on the Dutch
National Supercomputing cluster Cartesius and are funded by the NWO
computing grant 16431. The VLBA data shown in Figure \ref{fig:img-obs-data} 
is from project code: BW0106 PI: R. Craig Walker. This research has made use 
of NASA's Astrophysics Data System.\\

{\it Software:} {\tt BHAC} \citep{porth2017}, {\tt RAPTOR} \citep{bronzwaer2018}, {\tt eht-imaging} \citep{chael2016b,chael2018}, {\tt python} \citep{travis2007,jarrod2011}, {\tt scipy} \citep{jones2001}, {\tt numpy} \citep{walt2011}, {\tt matplotlib} \citep{hunter2007}, and {\tt rPICARD} \citep{Janssen2019}.

\bibliographystyle{aa}
%\bibliographystyle{stylename}
%\newpage
%\enlargethispage{5\baselineskip}
\bibliography{references}

\appendix

\section{Phenomenological model explaining the dominance of the counter-jet at 228 GHz}
\label{appendixA}

In certain GRMHD-based models of M87, when imaged at 228 GHz at low
($\sim$20 deg) inclination angles, it is observed that most of the
emission that reaches the observer originates from the `counter-jet',
i.e.~the jet facing away from the observer. Here we describe a simple
phenomenological model that is capable of reproducing this effect.

\begin{figure}[h]
  \centering
  \includegraphics[width=0.95\columnwidth]{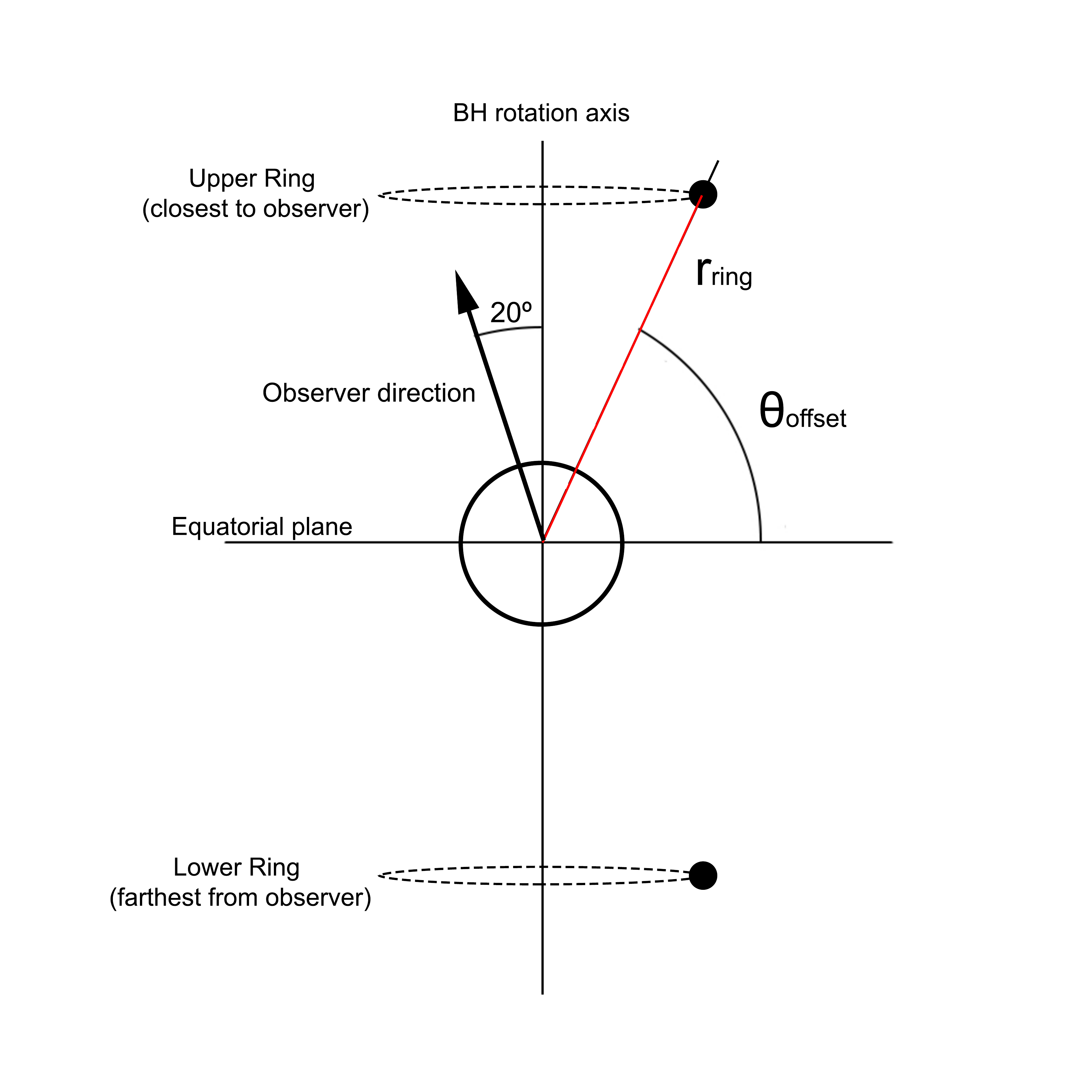}
  \caption{A diagram of our phenomenological model of the dominant
    counter-jet. The central circle represents the black-hole's event
    horizon. Two parameters control the placement of the rings (the model
    is symmetric with respect to the equatorial plane). The ring's
    thickness is set to 1 $~\rg$.}\label{fig:2ringmodel}
\end{figure}

Figure \ref{fig:2ringmodel} shows a schematic overview of our model,
which consists of two rings of luminous material. The model is symmetric
with respect to the equatorial plane, and the black-hole's rotation axis
passes through the center of the rings. The rings are meant to be an
approximation of the `jet base' which appears on both sides of the
equatorial plane in many GRMHD simulations. We make the assumption that
the luminous rings are perfectly optically thin (equivalently, we ignore
absorption in this model).

Two main parameters define the geometry of the rings: $\theta_{\rm
  offset}$ and $r_{\rm ring}$ determine their principal diameter and
distance from the equatorial plane. A step-function-based emissivity
profile is then used, which relates the emissivity at a location in space
to the distance between that location and the nearest point on the ring;
it is equal to unity if that distance is smaller than $0.25 ~\rg$, and
zero otherwise. In other words, the cross-sectional thickness of the ring
is $0.5 ~\rg$.

We also assign a velocity vector to the material in the ring; this is
done using a simple Keplerian model for the orbital velocity of a
particle with orbital radius $r_{\rm ring}$. The effect of this velocity
vector is to cause the characteristic relativistic-boosting effect seen
in most of our simulations; the ring is slightly brighter on the
approaching side. This effect is minor in the present case, due to our
low inclination angle.

\begin{figure}
  \centering
  \includegraphics[width=0.85\columnwidth]{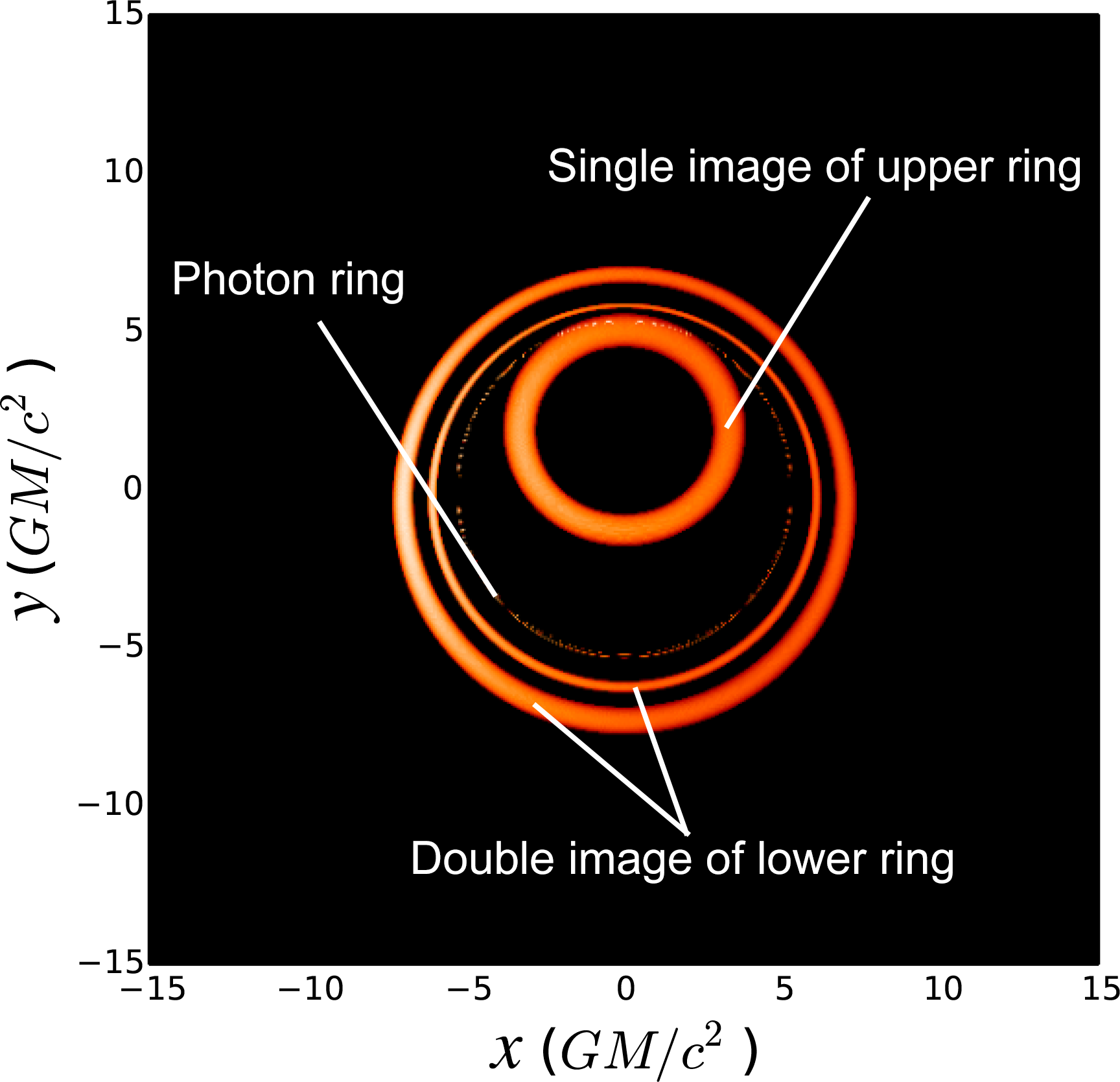}
  \caption{A typical image of our model, produced using {\tt RAPTOR} in
    the case of a Schwarzschild black hole ($a_*=0$). The observer
    inclination $i$ is 20 deg. $\theta_{\rm offset} = 1$ rad, $r_{\rm
      ring}=6 ~\rg$. Note the double image of the lower ring, which
    appears larger in size than the upper ring due to lensing. The
    doubled image of the lower ring appears close to the photon ring, but
    is slightly larger.}\label{fig:th_off_1_schwarz}
\end{figure}

\begin{figure}
\centering
\includegraphics[width=0.85\columnwidth]{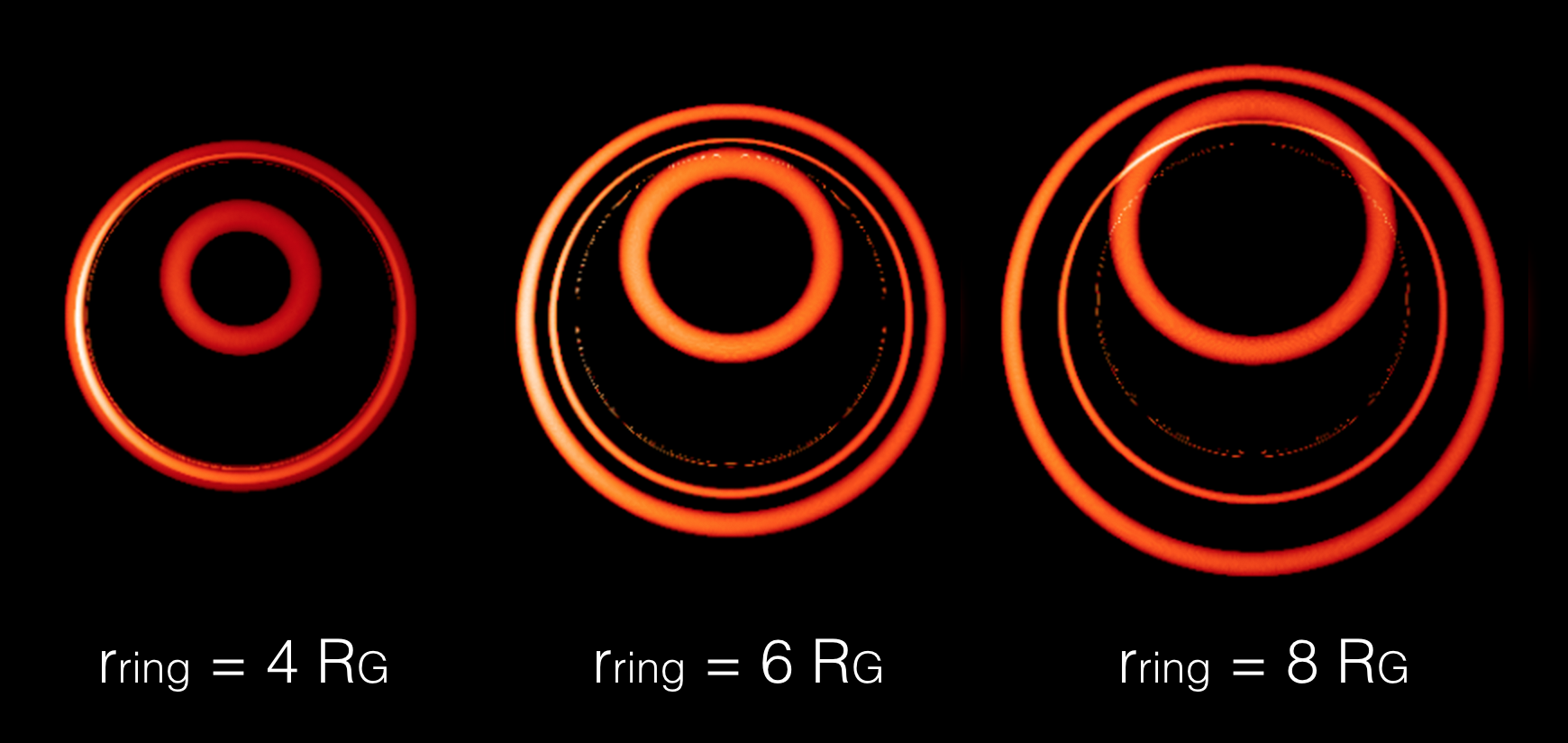}
\caption{Illustration of the effect of changing $r_{\rm ring}$, for a
  Schwarzschild black hole ($a_*=0$) imaged with $i=20$ deg and
  $\theta_{\rm offset} = 1$ rad. Note how the two images of the lower
  ring coincide for the case $r_{\rm ring} = 4
  ~\rg$.}\label{fig:diff_radii}
\end{figure}

\begin{figure}
\centering
\includegraphics[width=0.85\columnwidth]{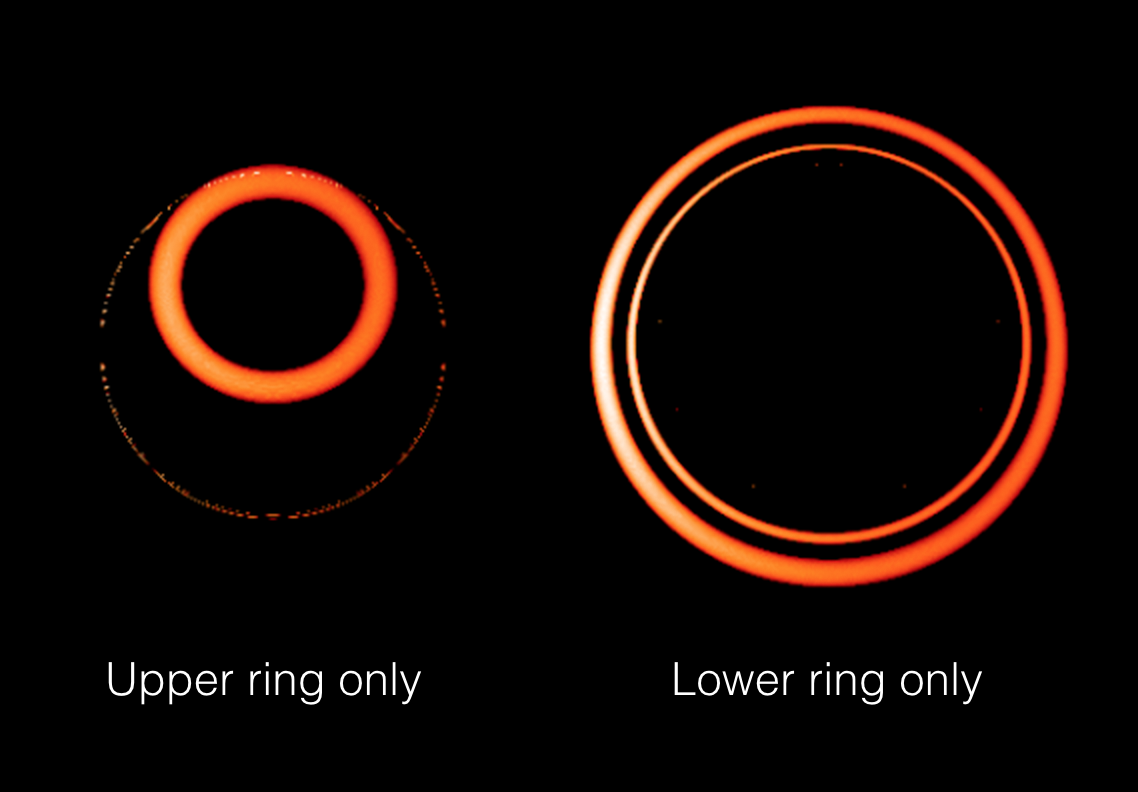}
\caption{Comparison of images which show only the upper (left) or lower
  (right) ring, omitting the other (as before, $a_*=0$, $i=20$ deg,
  $\theta_{\rm offset} = 1$ rad, $r_{\rm ring}=6 ~\rg$). The integrated
  flux density received from the lower ring is about 30\% higher than
  that from the upper ring, due to gravitational
  lensing.}\label{fig:upper_lower}
\end{figure}

Figure \ref{fig:th_off_1_schwarz} shows a typical image of our model,
with its key features annotated for clarity. Figure \ref{fig:diff_radii}
illustrates the effect of varying $r_{\rm ring}$. Figure
\ref{fig:upper_lower} compares two images that show only the upper
(lower) ring. The flux observed from the lower ring is 30\% higher than
that of the upper ring. Note that the lensed image(s) of the lower ring
always appear `outside' (but near) the black-hole's photon ring,
potentially causing the observer to overestimate the black-hole shadow
size. Although the black-hole mass can be derived from the size of the
black-hole shadow, such an estimate should be seen as an upper limit in
the present context.

In this model, gravitational lensing causes most of the radiation emitted
in the `polar regions' of a black hole to be redirected toward the
opposite side of the black hole with respect to its origin. This effect
could help to explain why the counter-jet in optically thin simulations
of M87 dominates over the observer-facing jet.

\begin{figure}[h]
\centering
\includegraphics[width=0.95\columnwidth]{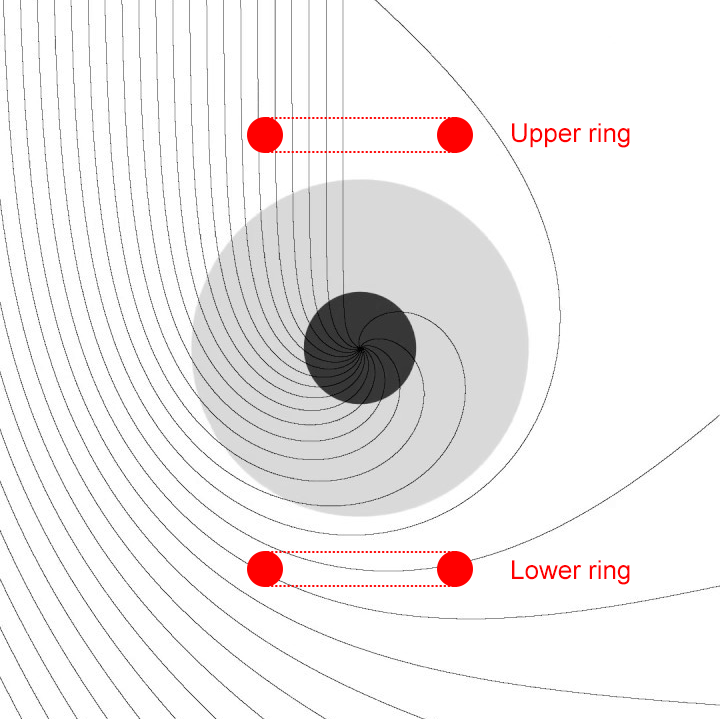}
\caption{Visualization of the `doubling effect'; the observer in this
  image is directly above the black hole (i.e., the inclination angle is
  zero). The black-hole's event horizon is marked by the dark-grey
  circle, while the unstable-photon region is marked by the light-grey
  circle. Gravitational lensing enhances the overall size of the lower
  ring, although the divergence of the rays near the lower ring causes
  its images to have a reduced thickness. Two images of the lower ring
  appear; one due to rays that intersect the ring while moving away from
  the observer, the other due to rays that curve around the black hole
  and intersect the ring while moving toward the observer. This causes
  most of the flux that reaches the observer to originate in the lower
  ring, on the far side of the black hole. Adapted from an image by
  Alessandro Roussel.}\label{fig:doubling}
\end{figure}

Figure \ref{fig:doubling} shows an illustration of the lensing effect
that causes an observer to record \emph{two} images of the lower ring,
and only one of the upper ring. We can understand this feature by
considering the rays shown in Fig.~\ref{fig:doubling} from left to right
(i.e., decreasing the rays' impact parameter): the deflection angle
increases as the rays curve more and more. The first object with which
the rays intersect is the lower ring - hence we see that as the widest
object on the observer's image. The next rays, curving even more,
intersect the lower ring again, but now they are moving back toward the
observer (having traveled around the black hole). Moving on to rays with
still smaller impact parameters, the rays now come very close to the
photon ring. At this point, the deflection angle begins to diverge,
causing rays to orbit the black hole an arbitrary number of times. These
rays image the entire sky infinitely many times, producing a multitude of
images of the environment. All of these images, however, are very small
(and thus they don't contribute much flux), and they are confined to a
thin ring, which is infinitesimally close to the photon ring.

Note that the doubling effect is only visible at low inclination angles,
when the system is viewed in a face-on manner; the effect vanishes
entirely at inclination angles near 90 degrees (symmetry demands that
both rings then contribute equally to the integrated flux density of the
image). Complications also arise when absorption is taken into account;
an optically thick accretion disk may absorb much of the lensed radiation
originating from the lower ring.

As a final comment, we note that the doubling of the lower ring due to
gravitational lensing occurs everywhere along the lower ring. Therefore,
a partial ring or even a very compact structure (e.g., a plasmoid or `Gaussian hot-spot') will show the same behavior; most of the radiation will
come from the hot-spot on the opposite side of the black hole, away from the
observer.

\end{document}